\documentclass{bmcart}

\usepackage{url}
\usepackage{hyperref}
\usepackage{amsthm,amsmath}
\usepackage{algorithm}
\usepackage[noend]{algpseudocode}
\usepackage[export]{adjustbox}

\newcommand{\DR}{\textsc{DR}}
\newcommand{\method}{\textbf{IMR}}

\usepackage[utf8]{inputenc} 

\startlocaldefs
\endlocaldefs

\begin{document}
	
	\begin{frontmatter}
		
		\begin{fmbox}
			\dochead{Research}
			
			
			\title{A Path-Based Approach to Analyzing the Global Liner Shipping Network}
			
			
			\author[
			addressref={aff1},
			email={larock.t@northeastern.edu}   
			]{\inits{T.L.}\fnm{Timothy} \snm{LaRock}}
			\author[
			addressref={aff2},
			corref={aff2},
			email={stephanie1996@sina.com}
			]{\inits{M.X.}\fnm{Mengqiao} \snm{Xu}}
			\author[
			addressref={aff1, aff3},                   
			email={eliassirad.t@northeastern.edu}   
			]{\inits{T.E.}\fnm{Tina} \snm{Eliassi-Rad}}
			
			
			\address[id=aff1]{
				\orgdiv{Network Science Institute},             
				\orgname{Northeastern University},          
				\city{Boston, MA},                              
				\cny{USA}                                    
			}
			\address[id=aff2]{%
				\orgdiv{School of Economics and Management},
				\orgname{Dalian University of Technology},
				\city{Dalian},
				\cny{China}
			}
			\address[id=aff3]{
				\orgdiv{Khoury College of Computer Sciences},             
				\orgname{Northeastern University},          
				\city{Boston, MA},                              
				\cny{USA}                                    
			}

			
			
		\end{fmbox}
		
		
		\begin{abstractbox}
			
			\begin{abstract} 
				The maritime shipping network is the backbone of global trade. Data about the movement of cargo through this network comes in various forms, from ship-level Automatic Identification System (AIS) data, to aggregated bilateral trade volume statistics. Multiple network representations of the shipping system can be derived from any one data source, each of which has advantages and disadvantages. In this work, we examine data in the form of liner shipping service routes, a list of walks through the port-to-port network aggregated from individual shipping companies by a large shipping logistics database. This data is inherently sequential, in that each route represents a sequence of ports called upon by a cargo ship. Previous work has analyzed this data without taking full advantage of the sequential information. Our contribution is to develop a path-based methodology for analyzing liner shipping service route data, computing navigational trajectories through the network that respect the routes and comparing these paths with those computed using other network representations of the same data. We further use these trajectories to re-analyze the role of a previously-identified structural core through the network, as well as to define and analyze a measure of betweenness centrality for nodes and edges.
			\end{abstract}
			
			
			
			
		\end{abstractbox}
		%
		
	\end{frontmatter}
	
	
	
	
	\section{Introduction}
	Maritime container shipping facilitates trade and logistics at the global scale. This global shipping system can be modeled as a complex network, with ports as nodes and edges between them representing flows of shipping vessels, containers, goods, or even invasive species \cite{hu2009empirical, kaluza2010complex, ducruet2010centrality, ducruet2012maritime, ducruet2012worldwide, ducruet2013network, xu2014improving, li2015centrality, xu2015evolution, kojaku2019multiscale, xu2020modular}. Analyzing this network can provide insight into factors important  to maritime economists and shipping industry experts, such as connectivity, efficiency, and robustness of the maritime shipping network, and thus the network of global trade. 
	
	The construction of the network, particularly the choice of connections between ports, depends first and foremost on the type of  data available. Many studies have used data from vessel tracking based on Automatic Identification System (AIS) data \cite{kaluza2010complex, xu2014improving} that provides fine-grained trajectories of individual vessels moving between ports. The availability and granularity of this data makes it a valuable resource, but it has a few drawbacks. First, it requires substantial effort to collect and clean. Second, AIS data does not retain information on the unique property of liner shipping: vessels move on fixed liner service routes (which generally include source and target ports with multiple intermediaries between these end-points). AIS data alone one cannot give precise information about which ports are on a same route, since a ship may be redeployed from one route to another \cite{xu2020modular}.
	
\begin{figure}
		\includegraphics[scale=0.7]{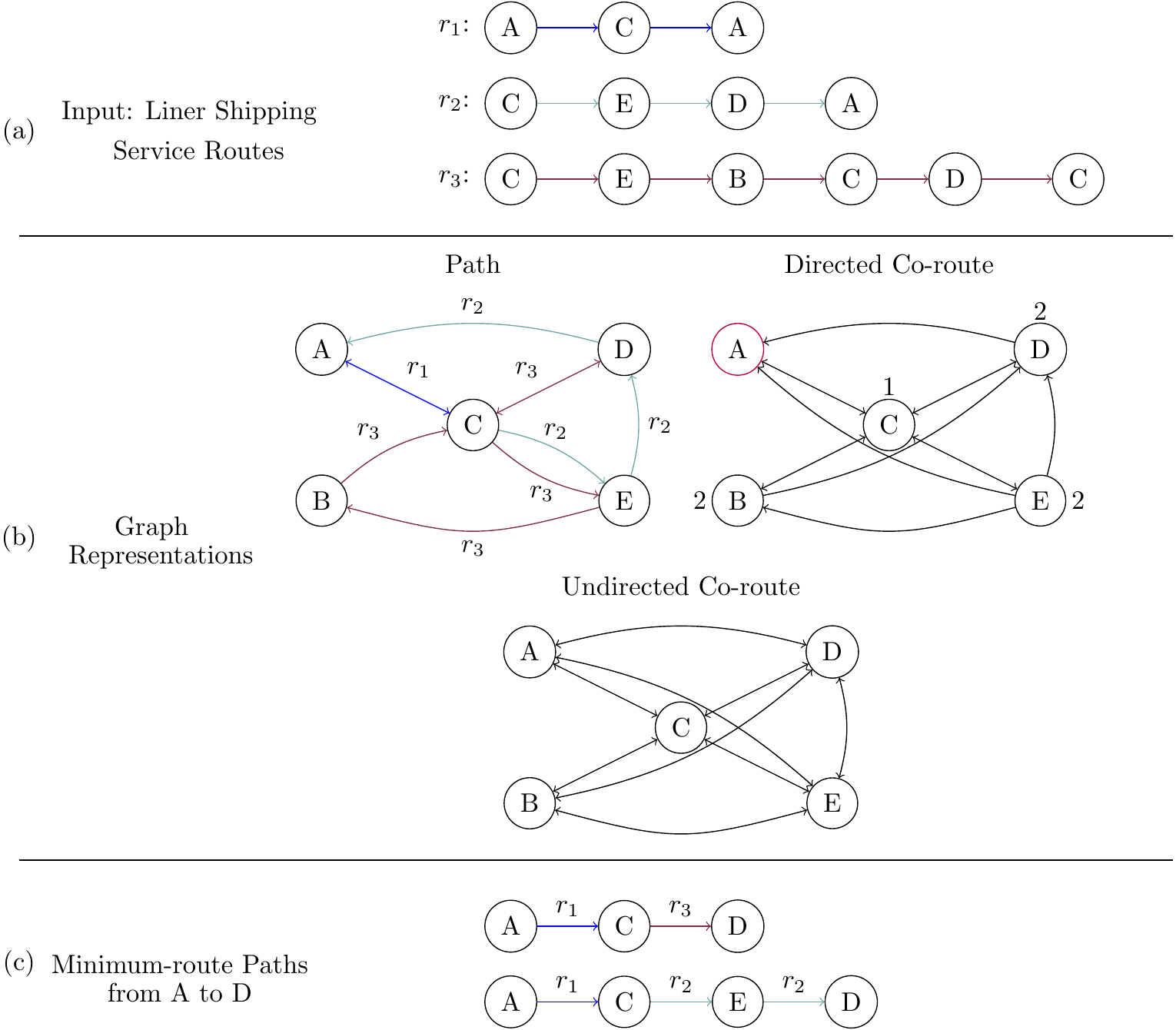}
		\caption{Examples of the three representations of liner shipping service route data studied in this paper. The input data is three routes (edge labeled) visiting the ports A, B, C, D, E. There are three routes labeled $r_1$, $r_2$, and $r_3$.The path graph is the traditional directed network representation of the routes, where an edge exists from u to v if the edge appears in at least one route. We also add parallel edges for every route in which the edge exists (or equivalently keep a set of route labels for each edge). This graph represents how ships and cargo can move through the network. The directed co-route graph is also a directed graph, but an edge exists from port u to port v if in at least one route port v appears in any succeeding ports of call after port u. The length of the shortest path between any two pairs of nodes in the co-route graph is the minimum-route distance (distances from A shown in (b)). In the undirected co-route graph, every route is made into a clique, or fully connected undirected graph. This representation was used for shipping route data in previous work \cite{xu2020modular}, emphasizing that cargo transportation between any two ports in a same route can be realized by one single vessel. All minimum-route paths between A and D, which require two routes and do not allow any port to appear more than once, are shown in (c).}
		\label{fig:explanatoryFigure}
	\end{figure}

	In this work, we analyze a different type of data: liner shipping service routes designed by container shipping companies and curated by Alphaliner, one of the largest proprietary shipping databases in the world.\footnote{Alphaliner: \url{https://www.alphaliner.com/}} The format of the data is a list of routes, where each route is an ordered sequence of ports called on by shipping vessels (see Figure~\ref{fig:explanatoryFigure}). Each of these routes can be conceptualized as a \emph{walk} through the port-to-port shipping network. Although this data is less granular than individual ship tracking with AIS data, it remains a valuable resource for understanding patterns in global container shipping \cite{xu2015evolution, xu2020modular}. Henceforth, we will refer to the data used in this study as the Alphaliner data. 
	
	Different network representations can be constructed from the Alphaliner data. Our work compares representations and methods used in previous analyses with new methods that are designed to account for sequential or pathway patterns inherent to shipping route data. Analysis of sequential patterns in complex networks is sometimes called \emph{higher-order network analysis}. Higher-order generally refers to interactions between nodes in a network that include more than two nodes at a time \cite{torres2021why}. These interactions may be unordered, as in hypergraphs \cite{chodrow2020configuration} and simplicial complexes \cite{battiston2020networksa}, or, as in our case, the interactions may be ordered, as in higher-order Markov models \cite{scholtes2017when, lambiotte2019networks}. The Alphaliner data is inherently sequential, since we know not only which ports are connected in dyads, but also which ports are visited as intermediaries between pairs of ports that do not have direct connections. The availability of this pathway information motivates the path-based approach we take in this study.
	
	Our goal is to develop a path-based methodology for liner shipping service route data that refines and expands our understanding of global container shipping. The Alphaliner data that we analyze, introduced fully in Section~\ref{sec:data}, is an aggregation of the liner shipping service routes on which shipping companies sent ships during the year 2015. Building on previous work that analyzed this data using complex networks \cite{xu2015evolution, xu2020modular}, our contribution is to interpolate and analyze the set of \emph{minimum-route paths} through the routes. A minimum-route path between ports $s$ and $t$ is a path that reaches $t$ starting from $s$ using the fewest number of \emph{transshipments} or \emph{transfers} between shipping routes. We compute and analyze these paths to improve our understanding of how cargo, specifically shipping containers, can move from one port to another through the port-to-port maritime shipping network. We choose minimum-route paths because they minimize the number of times a piece of cargo must be unloaded from one ship and loaded onto a ship traveling on a different route. Although we do not have data on specific container movements, the paths we compute represent plausible navigation trajectories for cargo moving through the global maritime shipping network based on the sequential information available in the liner service route data.
	
	In previous studies of liner shipping service route data a network was constructed by making each route into a fully connected undirected graph, then analyzing shortest paths through this network \cite{hu2009empirical, xu2020modular}. If the routes were bi-directional, this representation would have the advantage that the shortest path length between any two nodes is equal to the minimum number of routes required to move between them. This is not the case in the shipping route data used in these studies, making the path lengths through the network hard to interpret. Further, meaningful paths through the network are impossible to compute using this representation for two reasons. First, nodes that are only indirectly connected in a route are made to be directly connected. Second, all connections are made bi-directional, even though which directed edges exist can be discerned from the service routes themselves. These two issues make analyses that rely on shortest paths through this network representation potentially inaccurate.
	
	Our work addresses these inaccurate assumptions by incorporating route information into analysis of the global shipping network. We find a representation of the liner shipping service routes that contains realistic paths for cargo moving through the network of ports. We define and compute \emph{minimum-route paths} that reach a destination from a source using the minimum number of transfers between routes. We compare minimum-route paths with shortest paths through previously used network representations. Finally, we use minimum-route paths to define and evaluate \emph{route betweeness} measures of node and edge centrality in route data. We end with a discussion of the tradeoffs of the various representations, suggesting the sorts of analyses for which each representation may be best suited.
	
	The contributions of our papers are as follows:
	\begin{itemize}
		\item We compare three representations of shipping route data: the directed co-route graph, the undirected co-route graph, and the path graph (Section~\ref{sec:representations}).
		\item We provide an algorithm for computing \emph{minimum-route paths} from liner shipping service route data, called \method. These are paths that minimize the number of cargo transfers through the shipping network. We further provide algorithms for filtering minimum-route paths based on two factors, \emph{redundancy} and \emph{shipping distance}. We show that the properties of these paths differ substantially in terms of path length, shipping distance, number of routes required, and number of paths per pair of ports from the shortest paths used in previous analyses (Sections~\ref{sec:minroute-paths} \& \ref{sec:pathcomparison}). We further analyze the number of ways minimum-route paths may be realized and the relationship between the length of a liner shipping service route and the appearance of that route in minimum-route paths.
		\item We reanalyze the role of the \emph{structural core} of the global container shipping network defined in previous work \cite{xu2020modular}. We find that these analyses  \emph{underestimated} the role of core links in routing through the network for all paths between non-core ports and \emph{overestimated} their role in the subset of these paths that passed through the core (Section~\ref{sec:corecomparison}). We also find the role of local links was underestimated.
		\item We use minimum-route paths to define a modified betweenness centrality measure for both nodes and edges called \emph{route betweenness}. We analyze how this measure differs from topological centrality measures across representations (Sections~\ref{sec:methods-betw} \& \ref{sec:results-betw}).
	\end{itemize}
	
	\paragraph{Definitions} We note a few basic definitions that will come up repeatedly to avoid confusion. A \emph{walk} is a sequence of adjacent edges in a graph. Nodes and edges in a walk can be repeated. A walk is called \emph{closed} if it starts and ends at the same node and \emph{open} otherwise. A \emph{route} is a pre-defined walk through the shipping network. A \emph{path} is a walk that never repeats nodes, and a \emph{shortest path} between two nodes $s$ and $t$ is a path starting from $s$ and ending at $t$ visiting the fewest possible edges. Walks and paths can be directed or undirected, whereas in this work routes are always directed. We will also use the word path to refer to \emph{any} trajectory or sequence of edges in a network (e.g. the phrase path-based); we will note explicitly when we are using the term with its graph-theoretic meaning. We present a list of abbreviations and symbols we use throughout the paper in Table~\ref{tab:symbols} for reference. 
	
	{\renewcommand{\arraystretch}{1.3}
		\begin{table}[ht]
			\caption{List of Abbreviations \& Symbols}\label{tab:symbols}
			\begin{tabular}{|c|c|}
				\hline 
				Abbreviation & Meaning \\ 
				\hline 
				IMR & Iterative Minimum Route \\
				\hline 
				AIS & Automatic Identification System \\ 
				\hline 
				TEU & Twenty-foot Equivalent Unit \\ 
				\hline 
				$G_c=(V_c, E_c)$ & Directed co-route graph \\ 
				\hline 
				$R$ & Set of routes $r\in R$ traversing ports $p_i$, e.g. $r=<p_1, p_2, \cdots, p_\ell>$ \\ 
				\hline 
				$\ell_r$ & Length in edges of a route $r\in R$ \\ 
				\hline 
				$\textsc{MR}[d, s, t]$ & Minimum-route paths between ports $s$ and $t$ with distance $d$ \\ 
				\hline 
				$\textsc{DIST}[s, t]$ & Minimum-route distance between ports $s$ and $t$ \\ 
				\hline 
				$D_{\max}$ & Maximum minimum-route distance among all pairs \\ 
				\hline 
				$\eta_d$ & \# of pairs at minimum-route distance $d$ \\ 
				\hline
				$p_d$ & Max \# of minimum-route paths between any pair at distance $d$ \\ 
				\hline
				$\textbf{A}_{i,j}$ & Adjacency matrix of the path graph (unless otherwise specified) \\ 
				\hline
				PG & Path graph \\ 
				\hline 
				DCRG & Directed Co-route graph \\ 
				\hline 
				UCRG & Undirected Co-route graph \\ 
				\hline 
				$\ast$-betw & Betweenness in $\ast$ representation \\ 
				\hline 
				$\ast$-deg & Degree in $\ast$ representation \\ 
				\hline 
			\end{tabular} 
		\end{table}
	}

	\section{Related Work}
	We discuss some relevant work on the global liner shipping network, then discuss path problems in transportation networks more generally. We conclude this section by evaluating the position of our work relative to the studies discussed.
	
	\subsection{Global Liner Shipping Networks}
	Studies of the global liner shipping network can be categorized based on the type of data they analyzed (e.g. AIS, Port Authority, or shipping route databases), the methodology used (e.g. standard statistical analysis, optimization, or higher-order network analysis), and the particular aims of the research questions (e.g. economic analysis and modeling; analysis of structure, centrality, and hierarchy; or robustness and efficiency). While a full survey of these categories is outside the scope of this paper, in what follows we describe the research we believe is most pertinent to the present study.
	
	Kaluza et al. \cite{kaluza2010complex} constructed multiple port-to-port shipping networks from AIS data based on the type of cargo (bulk, oil, container). They analyzed properties of each of these networks, including distributions of (weighted) degree and clustering, the actual trajectories of ships through the network, and motif analysis. The study found that the global shipping network exhibits the small-world property and that there were substantial differences between how ships carrying different cargo navigated the network, suggesting implications for the spread of invasive species through ship ballasts. This latter result was built upon in \cite{xu2014improving, saebi2020network} (among others), where a network approach was again applied to AIS data to study the transfer of invasive species in ballasts of ships. The second study used techniques from higher-order network analysis, specifically the variable-order network model developed in \cite{xu2016representing}, to show that incorporating sequential information improves prediction of invasive species movements.
	
	The work most directly relevant to ours is the exploration of modular structure in the shipping network by Xu et al. \cite{xu2020modular}, which extended previous work by Hu et al. using liner shipping service route data \cite{hu2009empirical}. Hu et al. \cite{hu2009empirical} measured centrality and rich-club coefficients in maritime and air transportation networks, finding both to be organized hierarchically with rich-clubs. Xu et al. \cite{xu2020modular} identified a structural core in the global shipping network based on indices derived from partitioning the network into modules using community detection. They then analyzed the role of this structural core in navigating through the shipping network, finding that nodes and edges in the core play a substantial role in mediating or brokering movement through the network. Both of these studies used representations of the shipping route data we discuss at length below (Section~\ref{sec:representations}), the limitations of which we seek to address in this paper.
	
	\subsection{Paths in transportation networks}
	Sequential data is the basis for many studies of transportation networks, especially in public transportation. For example,  Barrett et al. presented an algorithm for solving the \emph{label-constrained shortest paths problem} in road and rail transportation networks, taking a formal languages approach \cite{barrett2008engineering}; Bast et al. proposed algorithms for solving time-constrained shortest-path problems in public transportation networks \cite{bast2010fast}; Lozano et al. presented a solution to the shortest viable path problem for multi-modal networks \cite{lozano2001shortest}; and Lewis et al. reviewed algorithms for computing shortest-paths with vertex transfer penalties \cite{lewis2020algorithms} (we do not have transfer times for our shipping routes).
	
	A closely related problem is finding walks through edge-colored graphs, for example the algorithms proposed in \cite{ferone2019kcolor}. However, typically the optimal solution is paths that use the maximum number of different colors, which in our case would correspond to using the largest number of unique routes, rather than the smallest.
	
	The work that comes closest to our own is \cite{bohmova2018computing}, which proposed algorithms for listing shortest paths in public transportation networks. However, the proposed algorithms assume that there are no cycles in the routes, e.g. that the routes are paths through the network, not walks. This assumption does not hold in the Alphaliner data. 
	
	Paths were constructed from public transit data for path-based analysis of the London Tube in \cite{scholtes2017when, larock2020hypa}. However, the method for constructing the paths was to compute shortest paths through the combined network of routes, which did not take the number of transfers into account.
	
	Although similar to much of the above work, our study differs on a few key points. First, many transportation systems, especially public transportation, evaluated in the previous studies are based on \emph{paths} through the network, since nodes are rarely if ever repeated in public transit routes. However, our shipping routes are not paths but \emph{walks}, since the same ports can be visited multiple times in a single route. Second, previous work has often (though not exclusively) focused on shortest paths, but our work will focus on minimizing the number of route transfers. 
	
	\section{Data}
	\label{sec:data}
	The Alphaliner data is given in the form of 1,622 liner shipping service routes from the year 2015, each of which is a sequence of $\ell$ ports visited by a ship on a single trip, e.g. $r=<p_1, p_2, \cdots, p_\ell>$. For each route, we are given an estimate of the total capacity of that route in 2015 in Twenty-foot Equivalent Units (TEU). Routes visit varying numbers of ports $\ell$ and we define the length of a route to be the number of legs $\ell-1$ it takes, finding that the average length of a route is 6.9 edges. The same port can be visited multiple times in a single route, making each route a \emph{walk} through the port-to-port network. A walk may be \emph{closed} (if the route starts and ends at the same port) or \emph{open} (if the route starts and ends at different ports). In this dataset, 1,416 of the routes are closed and 206 are open.
	
	\begin{figure}
		\centering
		\includegraphics[scale=0.4]{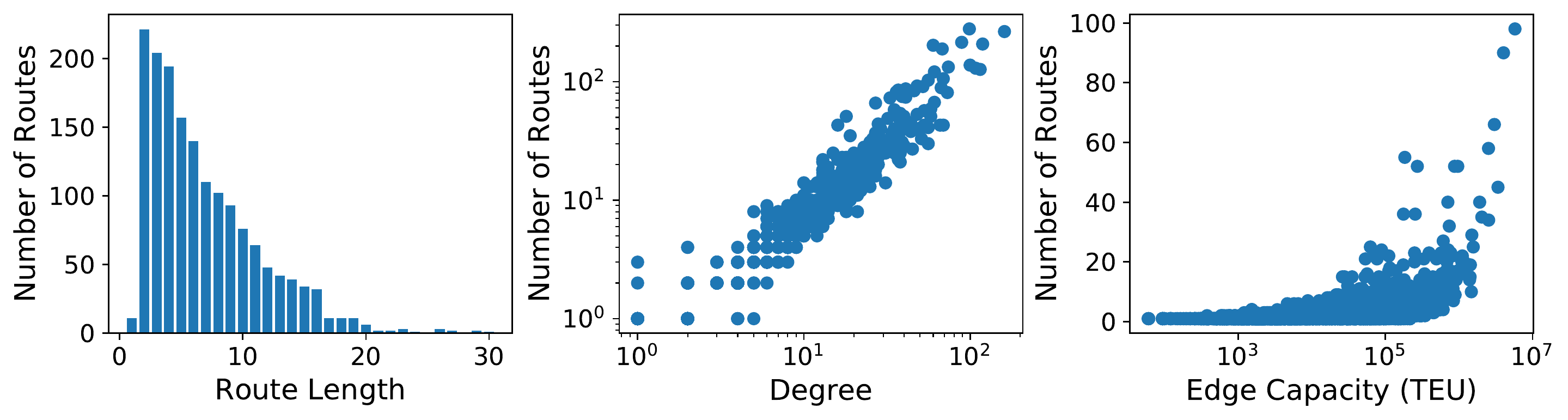}
		\caption{Statistics of the shipping routes. The left plot shows a histogram of the length of routes in terms of the number of edges. Most routes have length less than 10, but some visit as many as 30 ports. The middle plot shows the degree of a node (horizontal axis) against the number of routes the node appears in (vertical axis). The degree of a node in the path graph correlates with the number of routes it participates in and most nodes appear in a few routes, while some appear in more than 200. The right plot shows the estimated edge capacity in TEU (horizontal) against the number of routes in which each edge appears. Most edges appear in only a handful of routes, while few appear in nearly 100, and TEU capacity spans multiple orders of magnitude and correlates with the number of routes in which a port appears.}
		\label{fig:routestats}
	\end{figure}

	In Figure~\ref{fig:routestats} we show the distributions of route lengths (left), the degree of a node against the number of routes in which it participates (middle), and the estimated TEU capacity of an edge against the number of routes in which the edge participates (right). These distributions take similar forms, with the majority of routes being short and the number of routes each node and edge participates in being relatively small, but with tails in higher values. There are also positive correlations between the degree of a node and the TEU capacity of an edge with the number of routes in which the node or edge appears. This immediately suggests that ports and edges play different roles in the navigation of the network and that ports can be in principle separated into more central or important versus more peripheral, a common observation in shipping network analysis that is fundamental to the analyses that follow in this work \cite{ducruet2010centrality, xu2020modular}.
	
	\section{Background: Three representations of liner shipping service route data}
	\label{sec:representations}
	Previous work analyzing route data, including both maritime shipping and other transportation networks like public transit and railroad networks, have developed representations for the data that each have advantages and limitations. In this section we define three graph representations of route data and discuss their tradeoffs. We present examples of each representation in Figure~\ref{fig:explanatoryFigure} and statistics for each representation in Table~\ref{tab:representations}. All three representations include 977 ports, but the density in terms of number of edges, the average degree, and the average clustering differ substantially across the representations, as does the interpretation of the connections in the network.
	
	{\renewcommand{\arraystretch}{1.5}
		\begin{table}[H]
			\begin{tabular}{c|c|c|c|c|c}
				\hline
				Representation & Directed? & Nodes & Edges & Mean Degree & Mean Local Clustering \\ \hline
				Path & Yes & 977 & 5,268 & 5.4 & 0.26 \\ \hline
				Co-route (directed) & Yes & 977 & 30,035 & 30.7 & 0.64 \\ \hline
				Co-route (undirected) & No & 977 & 16,680 & 34.15 & 0.71 \\ \hline
			\end{tabular}
			\caption{Statistics for each graphical representation of the shipping route data. The path graph is substantially more sparse than the other two representations.}
			\label{tab:representations}
		\end{table}
	}
	
	\subsection{Path graph}
	The first representation, which we call the path graph, is the standard representation of a directed and weighted network. In the path graph, an edge exists between nodes $u$ and $v$ if that edge appears exactly in a route, e.g. there is some route $r$ such that $u$ and $v$ appear in sequence in $r$. In this representation we also label the edges with the routes in which they appear, equivalently formalized as many parallel labeled edges (1 per route the edge appears in), or a single set of routes as an edge attribute (with cardinality the total number of routes the edge appears in). The degree of a port $u$ in the path graph indicates the number of unique ports (excluding parallel edges) that can be reached directly from $u$. The weighted degree (or \emph{strength}) of a node $u$ is the total number of edges the node participates in across all routes (including parallel edges). This representation is the most sparse of the three we analyze, with 5,268 edges, average degree 5.4, and average local clustering coefficient 0.26.
	
	\subsection{Directed Co-route graph}
	In the directed co-route graph, a directed edge exists between $u$ and $v$ if there is some \emph{open} route $r$ such that $u$ appears before $v$ in $r$, or some \emph{closed} route $r$ such that $u$ and $v$ both appear in $r$ (the assumptions behind this construction are discussed further in Section~\ref{sec:minroute-paths}). In this graph, the shortest path distance between any two nodes $u$ and $v$ represents the minimum number of routes required to reach $v$ from $u$. However, shortest paths themselves through the graph may be misleading, since direct edges are drawn even where actual connections between ports are indirect. For example, in the co-route graph in Figure~\ref{fig:explanatoryFigure}, A-C-B is a shortest path between nodes A and C. However, the edge C-B never appears in the shipping routes. To get from A to B, a container would need to visit the edges C-E and E-B. In this representation, degree indicates the number of ports that can be reached through some other port, taking directionality into account. The strength of a node is its degree including all parallel edges representing different routes. This representation is more dense than the path graph, with 30,035 directed edges, average degree 30.7, and average local clustering coefficient 0.64.
	
	\subsection{Undirected co-route graph}
	In the undirected co-route graph representation each shipping route is made into a fully connected and undirected graph, i.e. a clique. If the routes being represented are bidirectional, then the shortest path length between nodes $u$ and $v$ in this graph reflects the minimum number of routes required to navigate between $u$ and $v$. However, the shipping routes used in this and previous studies are not bidirectional. This representation also suffers from the same problem as the co-route graph that shortest paths do not reflect actual navigation trajectories. For example, in the undirected co-route graph in Figure~\ref{fig:explanatoryFigure} there is a direct edge A-D, but this edge does not appear in the shipping routes. In this representation, degree indicates the number of ports that can be reached using a single route, assuming the routes have no directionality. This is the most dense representation, with 16,680 undirected edges (33,360 directed), average degree 34.15, and average local clustering 0.71.
	
	In the remainder of this work, we implicitly use the path graph representation as our network of interest, in contrast to some previous work on liner shipping service route data that studied shortest paths through the undirected co-route graph representation \cite{xu2020modular}. Rather than studying shortest paths, we compute paths that use the minimum number of route transfers, or \emph{mininmum-route} paths, which we define in the next section.
	
	\section{Proposed Methods}
	\label{sec:method}
	
	In this section we describe our methodology for studying the liner shipping service routes from a path-based perspective. We define minimum-route paths and a procedure, \method, for computing them, as well as additional procedures for filtering redundant and unrealistic paths. Then we use these paths to define measures of port and edge betweenness for route data.
	
	\subsection{Minimum-route Paths}
	\label{sec:minroute-paths}
	
	Intuitively, a \emph{minimum-route path} is a path from a source port to a target port that uses the minimum number of transfers between shipping routes. We are interested in these paths because they minimize the number of times a container needs to be unloaded and reloaded at a port, which is costly in terms of time, money, and coordination. In practice there are often many minimum-route paths between a given source and target pair. These paths are at least as long as the shortest path (in edges) between the source and target ports in the path graph, and may be longer if using the shortest path would require using more than the minimum number of routes. For example, if the path A-B-C-D connects A and D using only 1 route, while the path A-E-D connects the ports using 2 routes, only the first will count as minimum-route.
	
	Many shipping routes are closed walks, meaning they start and end at the same port. In industry practice, ships circulate on these routes regularly, meaning paths may continue from the end of the route on to the beginning. For example, in the route A-B-C-T-E-F-S-C-A, which starts and ends at the same port A, we consider the path S-C-A-B-C-T to be a valid minimum-route path between S and T that uses 1 route. Note that in these cases we allow cycles to occur in the route. There are alternative assumptions we could make that disallow cycles. We could use the path S-C-T, which only uses 1 route but also requires a transshipment, since a container traversing the path would need to be unloaded at port C, then loaded on another ship and brought to port T. This could be reasonable in some cases where the number of intermediate ports is very large, or it could be unreasonable if the number of intermediate ports is small. Alternatively, we could not allow any path from S to T using this route. This corresponds to a very strong assumption about directionality of the routes, but it misrepresents how the system operates, since routes are intentionally designed as closed walks.
	
	\subsubsection{Constructing Minimum-route Paths}
	We iteratively construct minimum-route paths directly from the shipping routes. Algorithm~\ref{alg:minroutes-iterative} contains pseudocode for our proposed procedure, \method. We are given as input the set of routes $R$. Using $R$, we construct the directed co-route graph representation of the routes $G_c=(V_c,E_c)$ where $V_c$ is the set of ports and an edge $(u,v)$ exists in $E_c$ if either (1) there is a closed route containing both $u$ and $v$, or (2) there is an open route such that $u$ appears before $v$. A port $t$ is reachable from a port $s$ if there is at least one path that follows the directed edges in $E_c$ from $s$ to $t$. Using $G_c$, we compute the set of all reachable pairs using Breadth First Search from every source node in $V_c$ and add each to the set of remaining pairs $P_R$. At the same time, we compute the shortest path distance $dist[s,t]$ for all pairs $(s,t)$ in the directed co-route graph, which is the same as the minimum-route distance. This allows us to identify which pairs have minimum-route paths at each distance. We use $D_{max}$ to denote the maximum minimum-route distance among all pairs of ports.
	
	We build the set of paths iteratively, starting with pairs that can be connected using 1 route. For this, we loop over each route $r \in R$, checking if the route is \emph{open}, meaning the first and last nodes are not the same, or \emph{closed}, meaning $r$ starts and ends at the same port. In the case where $r$ is open, we add all of the paths between each pair of indices $i,j, i<j$. If $r$ is closed, we add all paths between all pairs $i,j\in r, i\neq j$, allowing paths to continue from the end of the route to the beginning. Finally, we iterate over each minimum-route distance $d$, finding all minimum-route paths for pairs of nodes that require $d$ routes. At each distance, we loop over all pairs $(s,t)$ that are reachable using the current number of routes $d$. Then, we loop over all $d-1$-route paths from $s$ searching for any intermediate nodes $w$ that have a 1-route path to $t$ (by definition such a path exists). For any ports $w$ such that a path $s \cdots w \cdots t$ exists, we record all such paths. When minimum-route paths have been computed for all pairs at distance $d$, we restart the while loop until all pairs have been evaluated.
	
	\begin{algorithm}
		\caption{\textbf{IMR}($R$): Algorithm for computing \textbf{I}terative \textbf{M}inimum \textbf{R}outes}
		\label{alg:minroutes-iterative}
		\begin{algorithmic}[1]
			\Require $R$ (set of shipping routes)
			\Ensure $\textsc{mr}$ (minimum route paths)
			\State Construct directed co-route graph $G_c=(V_c, E_c)$ from routes $R$
			\State Compute shortest path distances $\textsc{dist}[s,t]$ in $G_c$ (BFS)
			\State $D_{\max} \gets \max_{s,t}\textsc{dist}[s,t]$
			\State Initialize minimum-route distance $d\gets 1$
			\Statex {\verb!// Compute all minimum-route paths that use exactly 1 route!}
			\For{all routes $r \in R$}
			\If{$r$ is open}
			\State Add all paths from $i$ to $j$ in $r$ s.t. $i<j$ to $\textsc{mr}[d,i,j]$  
			
			\Else
			\State Add all paths between all pairs $(i,j)\in r, i\neq j$ to $\textsc{mr}[d, i, j]$ 
			\EndIf
			\EndFor
			\Statex {\verb!// Compute paths for pairs with minimum-route distance! $d>1$}
			\For{$d \in 2,3,\dots,D_{\max}$}
			\For{pairs $(s,t)$ s.t. $\textsc{dist}[s,t] = d$}
			\For{$w \in \textsc{mr}[d-1, s]$ s.t. $t \in \textsc{mr}[1, w]$}
			\State Concatenate all paths from $s$ to $w$ with all paths from $w$ to $t$ and add to $\textsc{mr}[d,s,t]$
			\EndFor
			\EndFor
			
			\EndFor
		\end{algorithmic}
	\end{algorithm}
	
	\subsubsection{Runtime Analysis}
	\label{subsec:runtime}
	The runtime of the \method\ algorithm is the sum of the runtimes of (I) the construction of $G_c$ from $R$ (line 1), (II) the runtime of computing shortest path distances between all reachable pairs in $G_c$ (line 2), (III) the runtime of the first loop (lines 5-9), and (IV) the final for loop (lines 10-13). 
	
	Steps (I) and (III) can be computed together in one loop over the full set of routes $R$. We let $\ell_r$ be the length of a given route $r\in R$. Regardless of whether a route is open or closed, the operations that compute the minimum-route paths and add edges to $G_c$ require $O(\ell_r^2)$ time to process every pair of nodes in $r$. Therefore the running time is bounded by the number of routes $|R|$ multiplied by $\ell_{max}^2$, the length of the longest route squared, resulting in the worst case running time $O(|R|\ell_{max}^2)$.
	
	Step (II), computing shortest path distances between all reachable pairs in $G_c$, can be done in $O(|V_c|(|V_c|+|E_c|))$ by running Breadth First Search (BFS) from each node in $V$.
	
	Finally, step (IV) is the doubly nested for loops (lines 10-13). We defined $D_{\max}$ to be the maximum minimum-route distance among all pairs; for notational convenience we will refer to it as just $D$ here. We note that the worst case value of $D$ is the total number of routes $|R|$ (the case where all routes chained together in at least one ordering connect a pair of nodes that cannot be connected otherwise).\footnote{In Alphaliner dataset, $D$ is 8 while $|R|$ is 1,622.}
	
	We further define $\eta_d$ to be the number of pairs at minimum-route distance $d$ and $p_{d}$ to be the maximum number of minimum-route paths between any pair of ports at distance $d$. The maximum value of $\eta_d$ is $|V_c|^2-|V_c|$ in the case where all ports are mutually reachable at the same distance $d$. For example, given the route A-B-C-A-B, all pairs of nodes are mutually reachable in 1 route, meaning $\eta_1=3^2-3=6$ corresponding to pairs A-B, A-C, B-A, B-C, C-A, C-B. In fact this maximum can only be reached when $D=1$, since by definition edges can be navigated using exactly 1 route and so it is impossible for all ports to be connected at the same minimum-route distance $d>1$. Thus our upper bound on $\eta_d$ is loose when $d>1$.
	
	We also want an upper bound for the quantity $p_d$. An upper bound on the maximum number of minimum-route paths using $d$ routes between a pair of nodes is the maximum number of walks between any pair. Since walks through a graph can in principle contain an infinite number of cycles, we will use the fact that the set of routes $R$ is finite and compute a bound on the maximum number of walks between any pair of nodes using up to $d$ routes. For a given value of $d$, the loose upper bound we arrive at is the maximum value of the adjacency matrix $\textbf{A}$ of the path graph representing the routes raised to the sum of the lengths (in edges) of the $d$ longest routes $\ell_d$:
	
	$$\max_{i,j} \textbf{A}_{i,j}^{\ell_d}.$$
	
	This quantity represents the maximum number of paths between any pair that use the maximum number of edges among $d$ routes. We note that the distribution of route lengths has a tail in larger values (see Figure~\ref{fig:routestats}).\footnote{The minimum length of a route in the Alphaliner dataset is a single edge, the median length is 6 edges, mean length is 6.9 edges, and the maximum length is 30 edges. We further note that the number of edges used from a route in a minimum-route path is about half of the edges in the route on average (see Figure~\ref{fig:routeImportance}d).}
	
	Each iteration of the outerrmost for loop (line 10) involves $\eta_d$ iterations of the next for loop (line 11). In the worst case an iteration of the outer for loop requires $\eta_{d-1}$ iterations of the inner for loop, each of which takes worst case time $p_d p_1$, the maximum number of paths using $d$ routes times the number using 1 route. Thus the total running time for a particular value of $d$ is the product of these terms: $O(\eta_d\eta_{d-1}p_d p_1)$. An upper bound on this running time is the maximum of this time over all values of $d$ multiplied by the number of distances (iterations of the for loop in line 10)
	
	$$\max_{d\in1\dots D}{\eta_d\eta_{d-1}p_d p_1},$$
	
	which we can upper bound as
	
	$$ |V_c|^4  (\max_{i,j} \textbf{A}_{i,j}^{\ell_D})^2 $$
	
	Putting the three terms together, we have the running time 
	$$O\left(|R|\ell_{\max}^2 + |V_c|(|V_c|+|E_c|) +  D|V_c|^4( \max_{i,j} \textbf{A}_{i,j}^{\ell_D})^2 \right).$$ 
	
	When $D$ is 1, $\eta_D$ is equal to the total number of reachable pairs, meaning the second for loop will not be entered and the last term will be irrelevant. As $D$ grows toward its maximum $|R|$, the last term dominates the runtime. However, the upper bound approximation is worse at higher $D$, since the upper bound on $\eta_d$ is only tight at $D=1$, and the upper bound on $p_d$ weakens as $D$ grows because the approximation monotonically increases with $D$ (e.g. $\ell_{D+1} > \ell_D$ for all $D$ and so $A_{i,j}^{\ell_{D+1}} > A_{i,j}^{\ell_D}$) and is tightest for of a pair of nodes that is connected using all of the $D$ largest routes in $R$.\footnote{We have some evidence that this case is unlikely to appear often in real-world data. The longest minimum-route path in the Alphaliner dataset uses 81 edges and 4 routes, while $\ell_4=119$. Similarly, the longest path using $D=8$ routes is 44 edges, while $\ell_8= 228$.}
	
	\subsection{Filtering Minimum-route Paths}
	\begin{figure}
		\centering
		\includegraphics[width=0.98\textwidth]{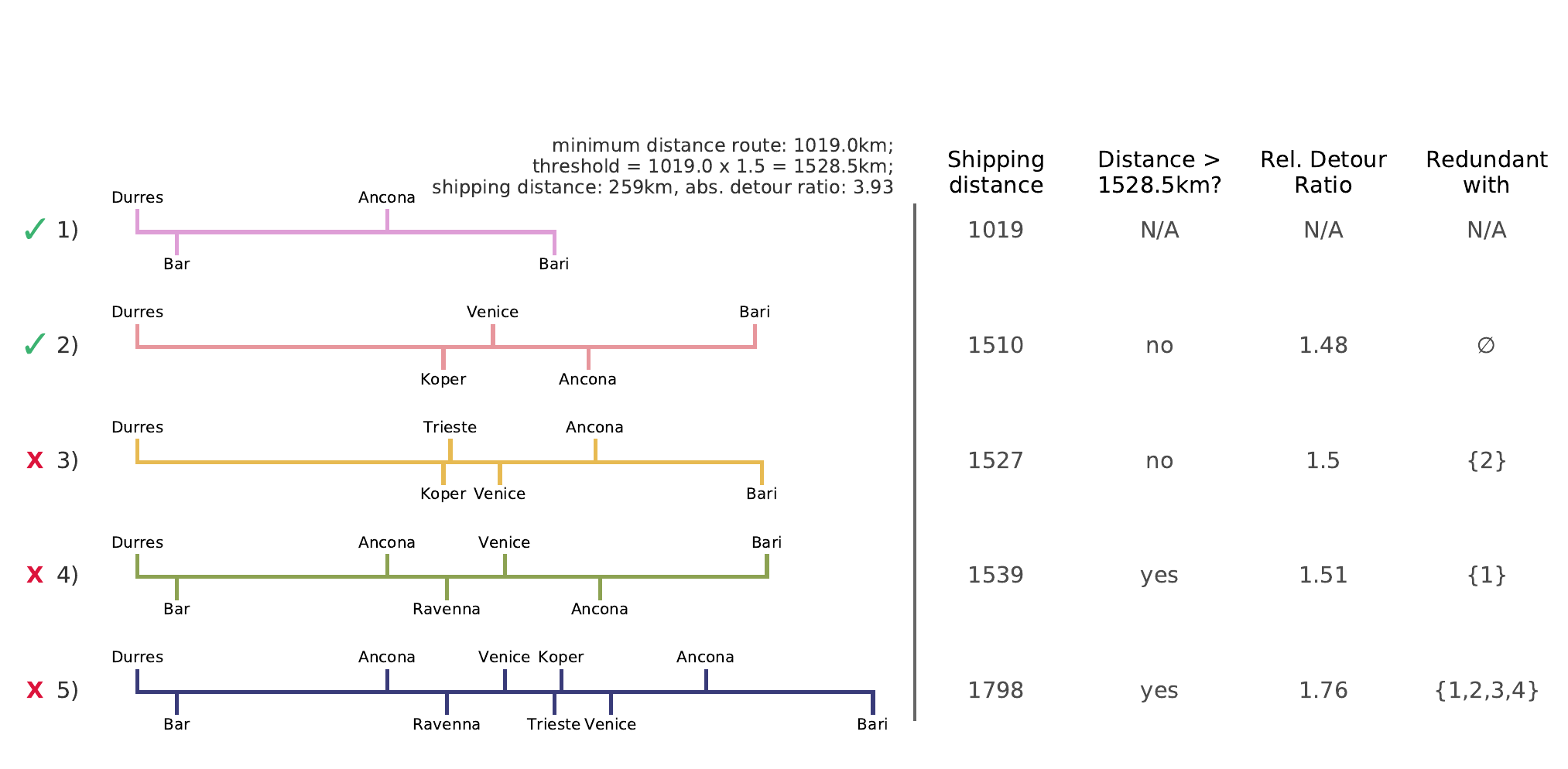}
		\caption{Example of our approach to filtering minimum-route paths. Paths 1 and 2 are kept because their shipping distances are not longer than 150\% of the shortest distance, nor are they redundant with any of the other paths. Paths 3, 4, and 5 are all redundant with at least one other path, and paths 4 and 5 are additionally filtered based on their shipping distance. In this example, none of the paths would be filtered based on the detour ratio method, since the absolute detour ratio (3.93) is larger than all of the relative detour ratios.}
		\label{fig:filteringex}
	\end{figure}

	There may be many paths between any given pair of nodes that use the minimum number of routes, and not all of these paths are equally desirable or plausible for navigation of the network. The assumption underlying the minimum-route paths is that shippers prefer to minimize trans-shipments, but other factors are also relevant to choosing between potential shipping routes. Two such factors are the total shipping distance of a route and the existence of shorter routes that visit essentially the same set of ports.
	
	\begin{algorithm}
		\caption{FilterPaths($\textsc{mr}, s, t, d, \alpha, \textsc{sd}$)}
		\label{alg:filter-paths}
		\begin{algorithmic}[1]
			\Require $\textsc{mr}$ (minimum route paths), $s$ (source port), $t$ (target port), $d$ (minimum-route distance), $\textsc{sd}$ (pairwise shipping distances)
			\Ensure $F$ (set of paths to filter)
			\State $F\gets \emptyset$
			\Statex {\verb!// Filter longer paths if they subsume any shorter paths!}
			\For{longer path $p_L$ in $\textsc{mr}[d, s, t]$}
			\For{shorter path $p_s$ in $\textsc{mr}[d, s, t]$ s.t. $|p_s|<|p_L|$}
			\If{$p_L\cap p_s \equiv p_s$}
			\State $F\gets F\cup p_L$
			\State \textbf{break}
			\EndIf
			\EndFor
			\Statex \hspace{0.49cm}{\verb!// Compute shipping distance for non-redundant paths!}
			\State $\textsc{dist}[p_L]\gets \sum_{i=1,\cdots,|p_L|-1} \textsc{sd}[p_L[i], p_L[i+1]]$
			\EndFor
			
			\State Filter distances using \textsc{ThresholdFilter}$(F, \alpha, \textsc{mr}[d,s,t])$ or \textsc{DetourFilter}$(F, \textsc{mr}[d,s,t])$
			\State Remove all paths $p \in F$ from $\textsc{mr}[d, s, t]$
		\end{algorithmic}
	\end{algorithm}
	
	Motivated by these considerations, we filter minimum-route paths using two criteria:
	\begin{enumerate}
		\item \textbf{Redundancy:} A path is excluded if there is a shorter path in which every port in the shorter path is also visited in the longer path. Concretely, given two paths $X$ and $Y$ with lengths $\ell_X>\ell_Y$, we say $X$ \emph{subsumes} $Y$ if $Y\subset X$. Any path that subsumes another path is called redundant and is filtered. For example, consider the longer path $X=A,B,C,D,E$ and the shorter path $Y=A,B,E$. The intersection between $X$ and $Y$ is all of $Y$, $X\cap Y\equiv Y$, meaning that $Y\subset X$ and $X$ is redundant. In contrast, the longer path $X'=A,F,G,E$ would \emph{not} be redundant with $Y$ because the node B does not appear in the longer path $X'$, meaning $Y\not\subset X'$ and thus $X'$ does not subsume $Y$. 
		\item \textbf{Distance:} We filter paths based on distance in two ways and compare the results. The first method uses a simple threshold on the shipping distance. A path is excluded if its total shipping distance is more than a factor $\alpha\geq 1$ of the minimum distance route. Concretely, for each pair $s,t$ we compute the total shipping distance (in km) for every path from $s$ to $t$, then set a threshold using the minimum of these distances multiplied by $\alpha$. Smaller $\alpha$ (closer to 1) will filter out more paths, since only paths with shipping distance close to the minimum will remain. If $\alpha = 1$ all paths except the minimum distance path are filtered, while $\alpha=\infty$ filters no paths. The second method uses the \emph{detour factor} of the minimum shipping distance path with the direct shipping distance and compares with the detour factor of all other paths with the minimum-distance path (described further in Section~\ref{subsec:detour}). We show pseudocode for both distance filtering methods in Algorithm~\ref{alg:filter-fncts}.
	\end{enumerate}
	
	\begin{algorithm}[ht]
		\caption{Procedures for distance filtering.}
		\label{alg:filter-fncts}
		\begin{algorithmic}[1]
			\Procedure{ThresholdFilter}{$F, \alpha, P=\textsc{mr}[d,s,t]$}
			\Statex \hspace{0.49cm}{\verb!// Filter paths with shipping distance longer than the minimum times!} $\alpha$
			\State $\textsc{dist}_{\min} \gets \min_{p\in P} \textsc{dist}[p]$
			\For{path $p \in \textsc{dist}$}
			\If{$\textsc{dist}[p] > \alpha\cdot \textsc{dist}_{\min}$}
			\State $F\gets F\cup p$
			\EndIf
			\EndFor
			\EndProcedure
			
			\Procedure{DetourFilter}{$F, P=\textsc{mr}[d,s,t]$}
			\Statex \hspace{0.49cm}{\verb!// Filter paths with relative detour factor larger than the minimum factor!}
			\State $\textsc{dist}_{\min} \gets \min_{p\in P} \textsc{dist}[p]$
			\State $\DR_{s,t}^{\min} \gets\frac{\textsc{dist}_{\min}}{\textsc{sd}[s, t]}$
			\For{path $p \in \textsc{dist}$}
			\If{$\frac{\textsc{dist}[p]}{\textsc{dist}_{\min}} > \DR_{s,t}^{\min}$}
			\State $F\gets F\cup p$
			\EndIf
			\EndFor
			\EndProcedure
		\end{algorithmic}
	\end{algorithm}
	
	We present pseudocode for our filtering procedure in Algorithm~\ref{alg:filter-paths}. The input to the algorithm is $mr$, the data structure output by Algorithm~\ref{alg:minroutes-iterative}; a pair of ports $s$ and $t$; the minimum-route distance between the ports $d$; the distance filtering threshold $\alpha$; and $sd$, a data structure containing the pairwise shipping distances between all ports. In the first outer loop we iterate over the paths $p_L$ from longest (in terms of edges) to shortest, then in the inner loop we iterate over all paths $p_s$ that are shorter than the current $p_L$. For each pair of paths, we check if $p_L\cap p_s \equiv p_s$, which indicates that the longer path subsumes the shorter path and thus should be marked redundant. If a path $p_L$ is not redundant, we compute and store in $\textrm{dist}[p_L]$ its total shipping distance as the sum of the distance between all adjacent ports in the path. We also compute the minimum distance in $\textsc{dist}_{\min}$. Finally, we filter the remaining paths based on distance in one of two ways presented in the next subsection.
	
	\subsubsection{Distance Threshold Filtering}
	In the first filtering method in Algorithm~\ref{alg:filter-fncts}, we compare the shipping distance of each path to this minimum shipping distance among all paths multiplied by the filtering threshold parameter $\alpha$, removing any paths $p$ such that $\textsc{dist}[p] > \alpha\textsc{dist}_{\min}$. We present an example using this method and analyze its runtime, then explain the parameterless detour factor filtering method.
	
	Figure~\ref{fig:filteringex} shows an example of the filtering process using  all of the minimum-route paths between the ports at Durress, Albania and Bari, Italy. The paths range in shipping distance from 1019km to 1798km, and some of the longer paths have significant overlap with shorter paths. The purpose of our filtering is to remove paths that are prohibitively long in terms of shipping distance (with respect to the minimum distance path between Durress and Bari) and those that are significantly redundant with shorter paths, since shippers are likely to simply choose the shorter of the paths. In the example, path 1 is automatically kept because it is the shortest both in terms of the number of edges used (3) and the total shipping distance (1019km). In this example, we set $\alpha=1.5$, meaning the shipping distance of a path must be less than 150\% of the minimum, in other words we set the threshold to be $1019 \times 1.5 = 1528.5$km. Based on this threshold path 2 is kept because its distance is less than 150\% longer than the first path and unlike path 1, path 2 does not visit the port at Bar. Path 3 is acceptable based only on distance, but it is redundant with path 2 because it visits all of the same ports, but adds a stop in Trieste, Italy. The final two paths are filtered because they are both too long (151\% and 176\% the minimum, respectively) and redundant with at least 1 other path. In fact, path 5 is redundant with all of the first four paths.
	
	\subsubsection{Path Filtering via Detour Factors}
	\label{subsec:detour}
	
	Using a threshold on the minimum shipping distance to filter out long paths has the advantage of simplicity, but the disadvantage that it is one size fits all, meaning the same thresholding factor $\alpha$ is used for every pair regardless of the distribution of shipping distances, and this parameter $\alpha$ must be set heuristically. This is unsatisfying because we expect these distributions to be different depending on the geographic distribution of the ports, with some quite far apart and others close together. An approach to filtering that does not require a single parameter to govern the filtering of all pairs of ports would be preferable. In this section, we develop such an approach using the \emph{detour factor} \cite{yang2018auniversal} (also known as the \emph{detour ratio}).
	
	Given two alternate paths $p_1$ and $p_2$ between ports $s$ and $t$ with respective (spatial) distances $d_1$ and $d_2$, we define the detour factor between the two paths to be $\DR(p_1, p_2) = \frac{d_1}{d_2}$. In the canonical detour factor, $d_2$ represents the great-circle distance between $s$ and $t$, guaranteeing that $d_1 \geq d_2$ and so $\DR(p_1, p_2) \geq 1$. 
	
	In this work we define two slight modifications of this usual definition. First, we define the \emph{minimum distance detour factor} $\DR_{s,t}^{\min}$ to be the detour factor when $p_1$ is the minimum shipping distance non-redundant path between $s$ and $t$ and $d_2$ represents the shipping distance (rather than the great-circle distance) between $s$ and $t$. Second, for each non-redundant path between $s$ and $t$ that is not the minimum shipping distance path, we define the \emph{relative detour factor} $\DR_{s,t}^{r_i}$ to be the detour factor when $p_1$ is the path in question and $p_2$ is the minimum shipping distance path between $s$ and $t$.
	
	Finally, we filter a path if its relative detour factor is larger than the minimum distance detour factor, e.g. if $\DR_{s,t}^{r_i} \geq \DR_{s,t}^{\min}$.
	
	\subsubsection{Runtime Analysis}
	\label{subsec:filt-runtime}
	
	In this section we analyze the runtime of the filtering procedure. Let $p_L$ represent the longest path (by edges) in $\textsc{MR}[d,s,t]$, and let $m=|\textsc{MR}[d, s, t])|$, the number of minimum-route paths between $s$ and $t$. The redundancy filtering dominates the computational complexity since it requires $O(m^2)$ time to loop over all $m$ paths. For both distance filtering methods, we need to compute the total distance for every path, which requires $O(|p_L|\cdot m)$ time in the worst case where all paths have the longest length. We can compute the minimum $\textsc{dist}_{\min}$ at the same time. Then we need to loop over the $m$ paths again to decide which need to be filtered. Therefore an upper bound on the running time is $O(m^2 + |p_L|\cdot m + m)=O(m^2 + |p_L|\cdot m)$. We observe that $m$ grows much faster than $p_L$ (see Figure ~\ref{fig:mVp-app} in Appendix~\ref{app:mVp}), thus in practice this running time is dominated by $O(m^2)$.
	
	The main factor in determining the running time for a specific pair of ports is the distribution of path lengths. If all paths have the same number of edges (which is unlikely), the redundancy computation can be skipped completely, since paths of the same length cannot be redundant. The more unique path lengths there are, and especially the more long paths that need to be compared with all shorter paths, the slower the computation will be. Further, the runtime of distance filtering is determined not only by the length of the longest path, but also by how many redundant paths are filtered before the distance filtering process begins, since these paths can be ignored.
	
	\subsection{Minimum-route Betweenness}
	\label{sec:methods-betw}
	Previous work has used betweenness centrality in the port network as a proxy for measuring the importance of a port to the navigation of the system \cite{ducruet2010centrality, xu2015evolution, xu2020modular}. Betweenness centrality for a node $u$ is defined as the sum of the proportion of shortest paths that include $u$ between each pair of nodes $(s,t)$ for all pairs $s\neq t \neq u$. We modify this definition by replacing the shortest paths between each pair with the set of (possibly filtered) minimum-route paths between $s$ and $t$. Using this alternative set of paths defines a measure we call \emph{route node betweenness centrality} that is based on navigation of the network using the shipping routes rather than shortest paths. Formally, route node betweenness centrality is computed as 
	
	$$ rb(u) = \sum_{s,t}{\frac{\sigma_{s,t}(u)}{\sigma_{s,t}}}$$
	
	where $\sigma_{s,t}(u)$ is the number of minimum-route paths from $s$ to $t$ that pass through the node $u$ and $\sigma_{s,t}$ is the total number of minimum-route paths between $s$ and $t$.
	
	We can also compute \emph{route edge betweenness centrality} following the same procedure as above except replacing nodes with edges:
	$$ rb(e) = \sum_{s,t}{\frac{\sigma_{s,t}(e)}{\sigma_{s,t}}}$$
	where $\sigma_{s,t}(e)$ is the number of minimum-route paths between $s$ and $t$ that use the edge $e$.
	
	\section{Experimental Results}
	In this section, we analyze the global liner shipping service route data using the path-based methodology set out in the previous section. We begin by evaluating the effect of the distance filtering parameter $\alpha$ on the minimum-route paths. Then, we compare four sets of paths constructed using different representations of the shipping route data: filtered minimum-route paths, shortest paths through the directed co-route graph, shortest paths through the undirected co-route graph, and shortest paths through the path graph. Next we build on previous work analyzing the \emph{structural core} of the global liner shipping network by comparing the role of core nodes and edges using minimum-route paths. Finally, we compare our measures of port and edge importance, route betweenness centrality, with external and topological measures of importance.
	
	\subsection{Filtered Minimum-route Paths}
	\begin{figure}
		\centering
		\includegraphics[scale=0.6]{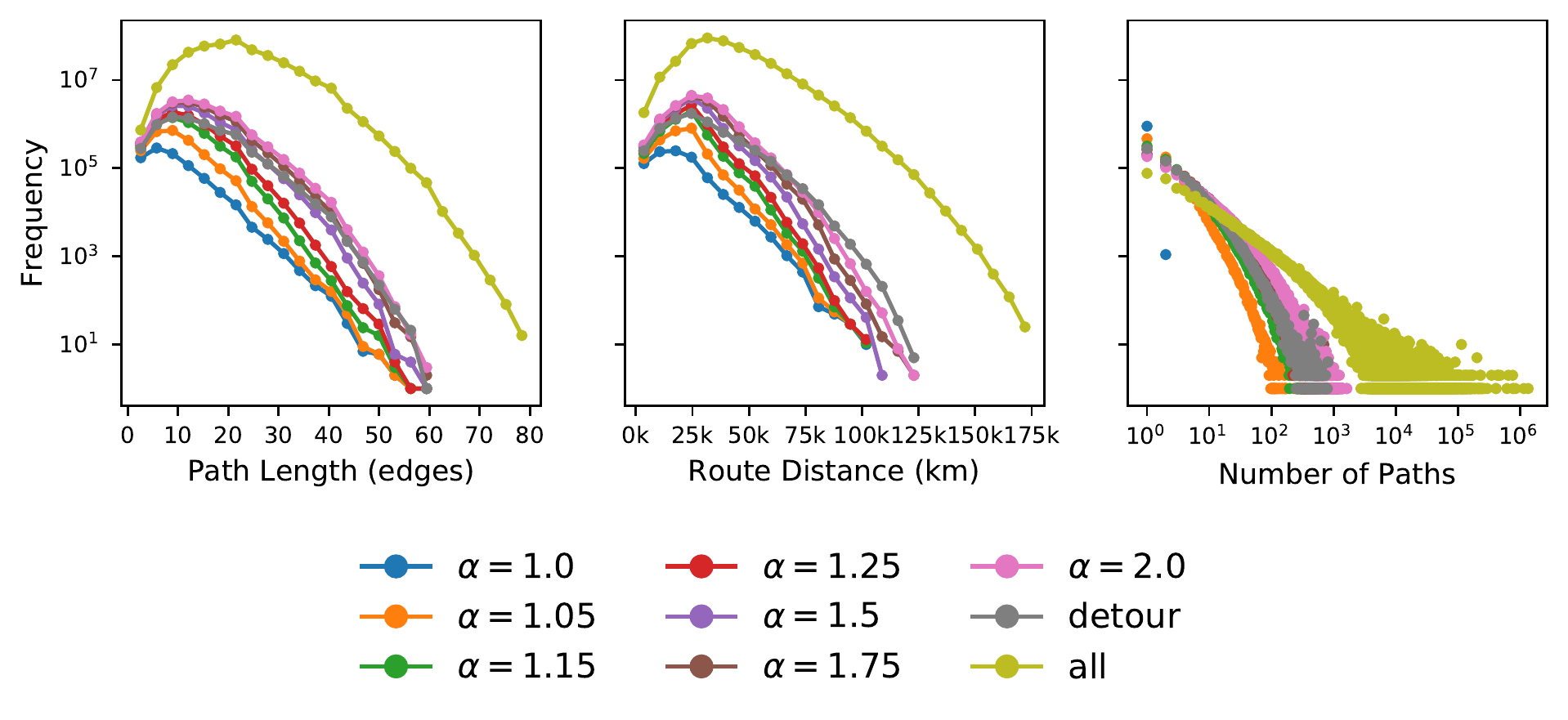}
		\caption{Outcome of filtering on minimum-route path statistics. Comparison of path length distribution (left), route shipping distance distribution (middle), and number of paths per pair (right) for all minimum-route paths against filtered paths with increasing values of the distance filtering parameter $\alpha$, as well as the detour rate thresholding method. All filtering methods reduce the maximum values in each distribution and more paths are filtered when $\alpha$ is closer to 1. The parameterless detour rate thresholding method is more conservative than filtering using $\alpha < 1.75$. We run subsequent analyses with $\alpha=1.15$, chosen heuristically because the results using this value fall somewhere in the middle in all distributions.
		}
		\label{fig:allVSfiltered}
	\end{figure}

	We compare path length and shipping route distance statistics for all minimum-route paths and filtered paths in Figure~\ref{fig:allVSfiltered}. We show results for $\alpha\in \{1.0, 1.05, 1.15,$ $1.25, 1.5, 1.75, 2.0\}$, as well as using the parameterless detour factor filtering. As $\alpha$ decreases, we see that long paths, both in terms of number of ports visited and total shipping distance, are reduced substantially. The number of paths per reachable pair is also reduced by multiple orders of magnitude after filtering, with the average number of paths per source and target pair dropping from 472 paths when including all minimum-route paths to an average of 6 paths after filtering with $\alpha=1.15$. Using the detour factor filtering behaves similarly to filtering with the largest threshold we tested, $\alpha=2.0$. We attribute the looseness of the filtering to the very large absolute detour factor between for some pairs of ports, which make it unlikely that the relative detour factors for the other paths will be large enough to filter. 
	
	Throughout the remainder of the analysis we use filtering threshold $\alpha=1.15$ unless otherwise indicated. We choose this threshold because it strikes a balance between short maximum path and route lengths and filtering out almost all paths.
	
	\subsection{Comparing Minimum-route Paths with Shortest Paths}
	\label{sec:pathcomparison}
	\begin{figure}[!ht]
		\centering
		\includegraphics[width=0.95\textwidth]{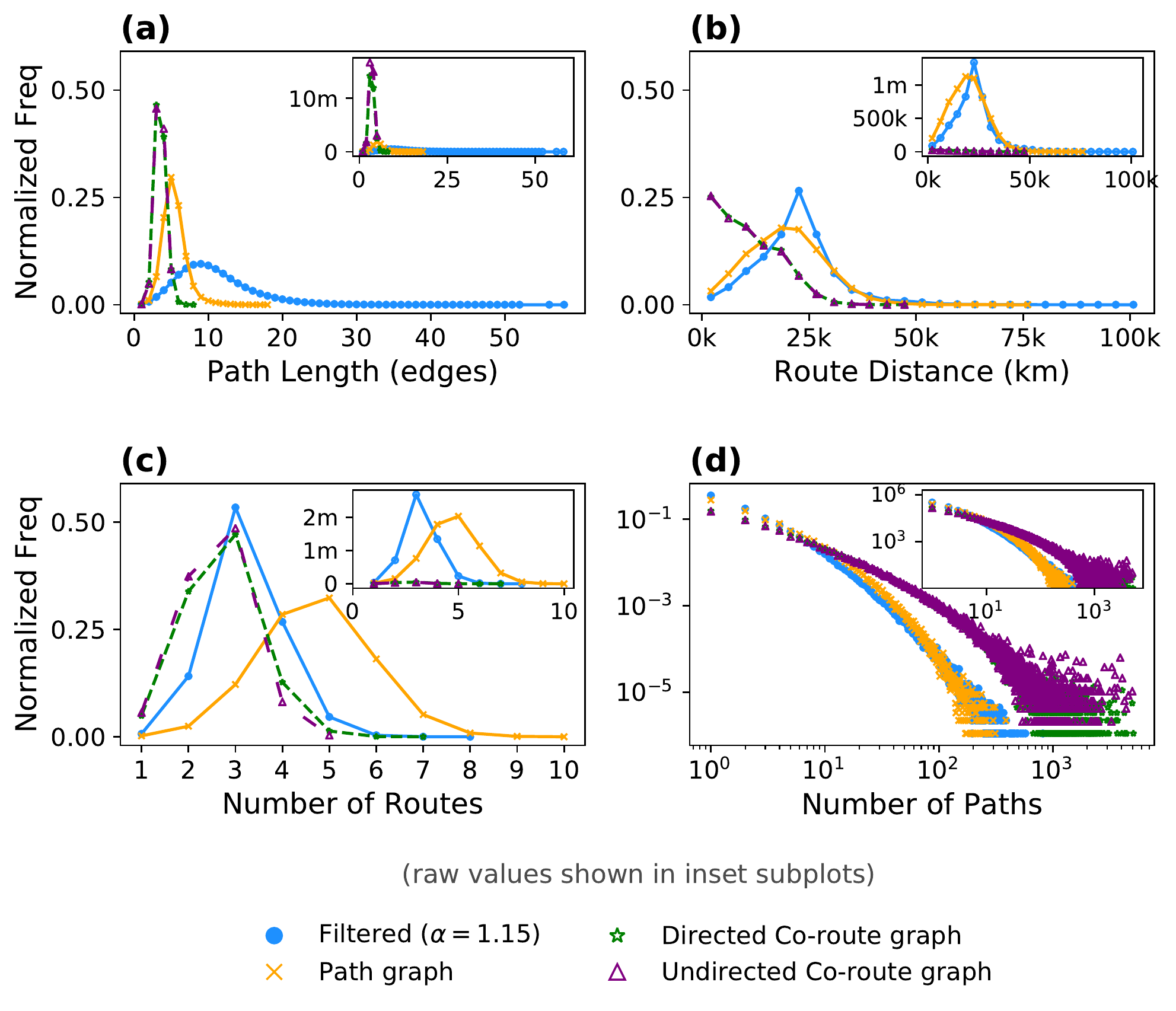}
		\caption{Filtered minimum-route paths more accurately capture how a container can move through the network based on liner shipping service routes. Path and route length statistics differ substantially between the filtered minimum-route paths (blue circles) and shortest paths through all of the path graph (orange crosses), the directed co-route graph (green stars), and  the undirected co-route graph (purple triangles). Each panel shows a normalized histogram with an unnormalized version in the inset plot. Network path lengths (a): Minimum-route paths are much longer (mean 20 edges) than the shortest paths through the undirected (3.5) or directed (3.5) co-route graph and the path graph (mean 5.4). The distribution of shipping route distances (b) for the minimum-route paths (mean 21,861.3 km) is longer than for the undirected (10,329.5) and directed (10,342.6) co-route graph and path graph (mean 19,532.4). The number of routes required (c) for the filtered paths (mean 3.2 routes) is shorter than in the path graph (4.7), where each shortest path may use a different number of routes. Only 0.3\% of the shortest paths through the undirected and 0.4\% of paths through the directed co-route graph are viable based on the shipping routes, but those that do correspond to viable paths use the minimum number of routes (means 2.6 and 2.7, respectively). Finally, the number of paths per pair of ports (d) is reduced substantially in the path graph and filtered minimum-route paths.}
		\label{fig:filteredVScgVSpg}
	\end{figure}

	In this section we compare the filtered minimum-route paths described above with shortest paths through the co-route graph, shortest paths through the undirected co-route graph (used in \cite{xu2020modular}), and shortest-paths through the path graph without using route information.
	
	Since the undirected co-route graph is undirected and includes many direct connections that are only indirect in the other two representations, it is not possible to make an exact comparison. Out of the 900,190 pairs of ports for which we have computed minimum-route paths, only 85,781 of those pairs have at least one shortest path through the undirected co-route graph that is also viable in the path graph.
	There is also a mismatch in the opposite direction: since many more ports are mutually reachable in the undirected co-route graph, there are 953,552 pairs of ports with at least one shortest path between them, more than the number of pairs that are connected by minimum-route paths. 
	
	This lack of alignment is impetus to take some care in explaining how we compute and report the distributions in Figure~\ref{fig:filteredVScgVSpg}. To make the potential issues concrete, there could be five shortest paths through the undirected co-route graph for a given pair. Each of these paths has the same length (by definition of the shortest path), but they may each use a different number of routes, and some may not be viable at all based on the routes. On the other hand, there may be ten minimum-route paths between the same pair, all of which use the same number of routes, but each of which is a different length. For the sake of comparison, in Figure~\ref{fig:filteredVScgVSpg}, where we compare distributions of path length, shipping route distance, and number of routes used across the three sets of paths, we (1) plot both normalized histograms (main plots) to compare the distributions directly and unnormalized histograms (inset) to get a sense for the differences in scale; and (2) compute one value per path between every pair, even when the values are all the same between that pair (in this sense we may describe the histograms as weighted). 
	
	{\renewcommand{\arraystretch}{1.3}
		\begin{table}
			\begin{tabular}{l|l|l|l|l|l}
				\hline
				Paths & \# pairs & \# paths & Avg paths / pair & Avg edges & Avg routes \\ \hline
				Path graph shortest & 900,190 & 6,289,093 & 7.0 & 5.4 & 4.7 \\ \hline
				Co-route (dir) shortest & 900,190 & 30,537,099 & 33.9 & 3.5 & 2.7 \\ \hline
				Co-route (undir) shortest & 953,552 & 36,565,016 & 38.3 & 3.5 & 2.3 \\ \hline
				Minimum-route & 900,190 & 424,919,483 & 472.0 & 20.0 & 3.7 \\ \hline
				Minimum-route ($\alpha=1.15$) & 900,190 & 5,034,897 & 5.6 & 10.7 & 3.2 \\ \hline
			\end{tabular}
			\caption{Path statistics for shortest paths through the path, directed and undirected co-route graphs, and minimum-route paths with and without filtering.}
		\end{table}
	}
	
	In the distribution of path lengths based on the number of edges used (Figure~\ref{fig:filteredVScgVSpg}a), we see that the shortest paths through the directed and undirected co-route, and path graphs are short compared to the filtered minimum-route paths. This is true both in terms of the average and maximum path lengths. This is important because it suggests that using shortest paths in analyzing this dataset will significantly underestimate the number of ports required for cargo to move between ports.
	
	In Figure~\ref{fig:filteredVScgVSpg}b we compare the distributions of route distance in kilometers for each set of paths. Most paths through the co-route and undirected co-route graphs are not viable in the routes and we do not compute distances for non-viable paths, however most viable paths have relatively short distances, with a maximum around 50,000km. The distance distributions for the minimum-routes and path graph paths have similar peaks between 19,000km and 24,000km, but the tail is shorter in the path graph paths. This again suggests that using shortest paths without accounting for sequential patterns in the routes may underestimate the distance required to ship a container.
	
	We compare the distributions of routes used per path in Figure~\ref{fig:filteredVScgVSpg}c. As expected, the minimum-route paths use fewer routes than the shortest paths through the path graph. We note that in the case where a shortest path through the directed or undirected co-route graphs does correspond to a viable path through the routes, we know that path is minimum-route because each step away from the source port in these representations corresponds to the use of 1 more route \cite{hu2009empirical}.
	
	Finally, in Figure~\ref{fig:filteredVScgVSpg}d, we compare the distributions of number of paths per pair of reachable nodes. Some reachable pairs in the undirected co-route graph have upwards of 1000 shortest paths between them, while the maximum number of paths for the path graph shortest paths and minimum-route paths is an order of magnitude less.

	\subsection{Route Importance}

	\begin{figure}
		\centering
		\includegraphics[width=0.95\textwidth]{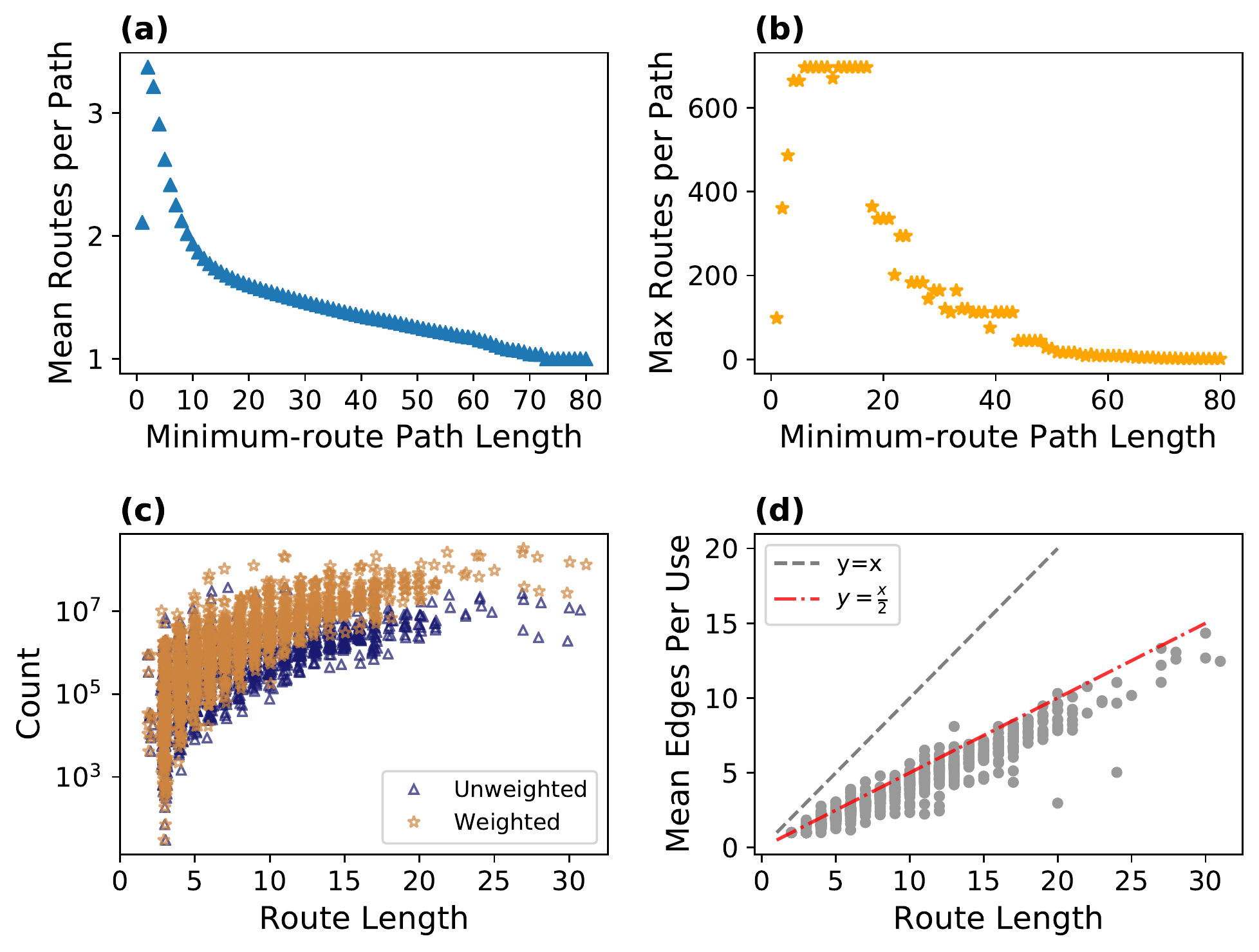}
		\caption{Many route sequences can realize the same path, and the amount of edges used from a given route is proportional to its length on average. The first two panels show minimum-route path length (horizontal) against the (a) mean and (b) maximum number of route sequences for minimum-route paths of that length. The mean number of route sequences per path is low (less than 4) across all path lengths, but maximums reach as high as 700 sequences realizing a single path. Panel (c) shows the length of a shipping service route (horizontal axis) against the number of route sequences that route appears in (unweighted, blue triangles), and the total number of edges from that route used across all minimum-route paths (weighted, orange stars). Longer service routes tend to be more highly weighted, but some shorter routes are also prevalent in realizing minimum-route paths. Panel (d) shows the average number of edges used per appearance in a route sequence for each service route (weighted divided by unweighted counts). The length of a route is a natural limit on the number of edges that can be used from that route in a given path; if the points were to fall along the line $y=x$, then each time a route appeared in a route sequence we could expect most or all of its edges to be used in that minimum-route path. Instead we find that on average roughly half of the edges from a route are used whenever it appears in a route sequence ($R^2=0.92$).}
		\label{fig:routeImportance}
	\end{figure}

	For each minimum-route path we compute, we also store every \emph{route sequence} that can realize the path. Since many of the same edges appear in more than one route, a minimum-route path can often be realized using multiple unique sequences of routes, each of which is a route sequence. Consider the simple case of the edge C-E in Figure~\ref{fig:explanatoryFigure}: the edge appears in both of the routes $r_2$ and $r_3$. Now consider adding a fourth path $r_4=E\rightarrow F$. Now to get from C to F, we can use two unique route sequences: $r_2, r_4$ and $r_3, r_4$. In this section we analyze these route sequences. 
	
	We begin by analyzing the number of route sequences per minimum-route path. The first two panels of Figure~\ref{fig:routeImportance} show the mean (a) and maximum (b) number of route sequences per minimum-route path length. For every minimum-route path length, the average number of route sequences per path is relatively low, less than 4. However, the maximum number of route sequences is upwards of 600 for many lengths between 5 and 20, and over 100 for paths as long as 40 ports. This indicates that some paths are realizable using numerous combinations of routes, which suggests these paths may be of particular importance to routing of cargo through the network.
	
	We now turn to the counts of the routes within the route sequences. For every minimum-route path, we loop over all route sequences that can realize that path. For each sequence, we increment two counters: an \emph{unweighted} counter is incremented by 1 for each route that appears in the sequence, and a \emph{weighted} counter is incremented by the total number of edges that are used from each route. The unweighted count corresponds to the number of unique route sequences that include the route, while the weighted count corresponds to the total number of edges from a route used across all minimum-route path realizations. We show histograms of these counts in Figure~\ref{fig:routeImportance}b. We observe that longer routes indeed tend to appear most often, however not all of the highly weighted routes are long, some appear to be shorter than 10 ports.
	
	Given that the shipping service routes and minimum-route paths both have varying lengths, we are interested in understanding the relationship between the length of a route and how much it appears in the minimum-route paths. In Figure~\ref{fig:routeImportance}c, we show the length of a route against the average number of edges used when that route appears in a route sequence, computed as the weighted count of the route divided by the unweighted count. We find a correlation between the length of a route and the average number of edges. This is to be expected, since there is a natural limit on the number of edges that can be used from short routes, e.g. a maximum of 1 edge can be used from a route of length 1. However, the maximum correlation in this case would be equality, meaning the whole service route was used every time it appeared in a minimum-route path realization. In reality, the average edges per use is about 50\% for the longest routes (the simple model $y=\frac{x}{2}$ shown in Figure~\ref{fig:routeImportance}d has coefficient of determination $R^2=0.92$).

	\subsection{Structural Core Analysis}
	\label{sec:corecomparison}
	
	Previous work classified connections between ports in the global shipping network into three categories based on whether ports involved in the connections were part of the “structural core” of ports, finding that this core plays an important role in supporting cargo transportation between non-core ports \cite{xu2020modular}. Using the undirected co-route graph representation, the structural core was defined by first computing a partitioning of the nodes in the network based on modularity-maximizing community detection (using the Louvain algorithm \cite{blondel2008fast}). The structural core of the graph was chosen to be a set of nodes such that (1) at least one node from each of the modules was present and (2) the density of connections among the nodes in this core was relatively high. A specific set of nodes was found that satisfied these criteria (using a heuristic choice of 0.8 subgraph density): the top 37 nodes with highest value of \emph{Gateway-ness}, a measure of the extent to which a port was connected to other ports outside of its own module, for a specific partition of the network. In this section, we compare the original analysis of the role of this structural core with an analysis using minimum-route paths rather than shortest paths through the undirected co-route graph.
	
	With the ports making up the structural core defined, we continue following \cite{xu2020modular} by categorizing each edge based on whether the ports on either end are in the structural core. The original taxonomy had three categories: \emph{core edges}, when both ports are in the structural core; \emph{feeder edges}, when exactly one port in the core; and \emph{local edges}, when neither port is in the core. Since minimum-rout paths take directionality into account, we can split the edges in the feeder category into two categories, where an edge is an \emph{out-feeder} if it points from a core node to a non-core node, and an \emph{in-feeder} if it points from a non-core node to a core node.

	\begin{figure}
		\centering
		\includegraphics[width=0.47\textwidth]{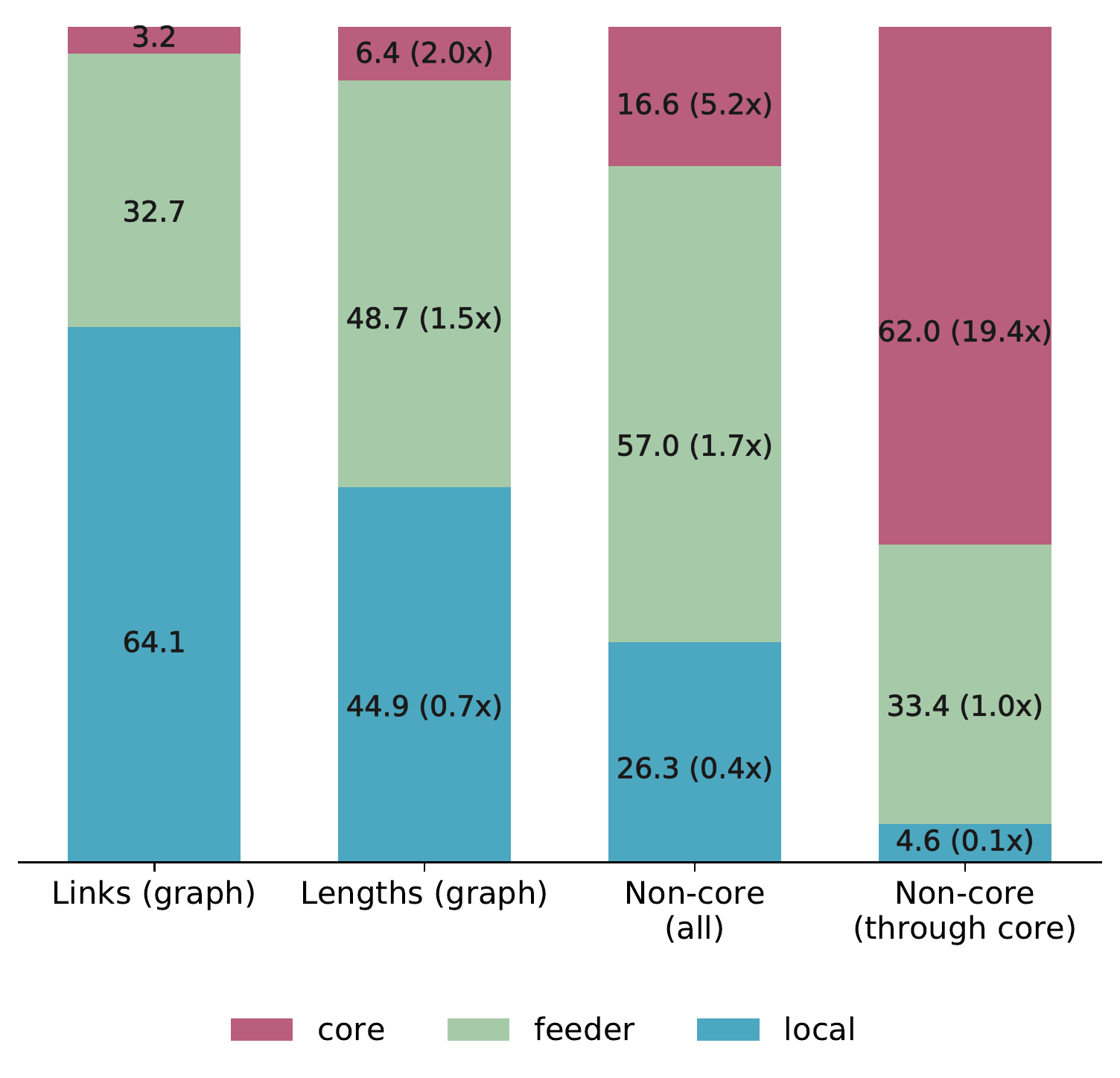}\hspace{0.25cm}\includegraphics[width=0.47\textwidth]{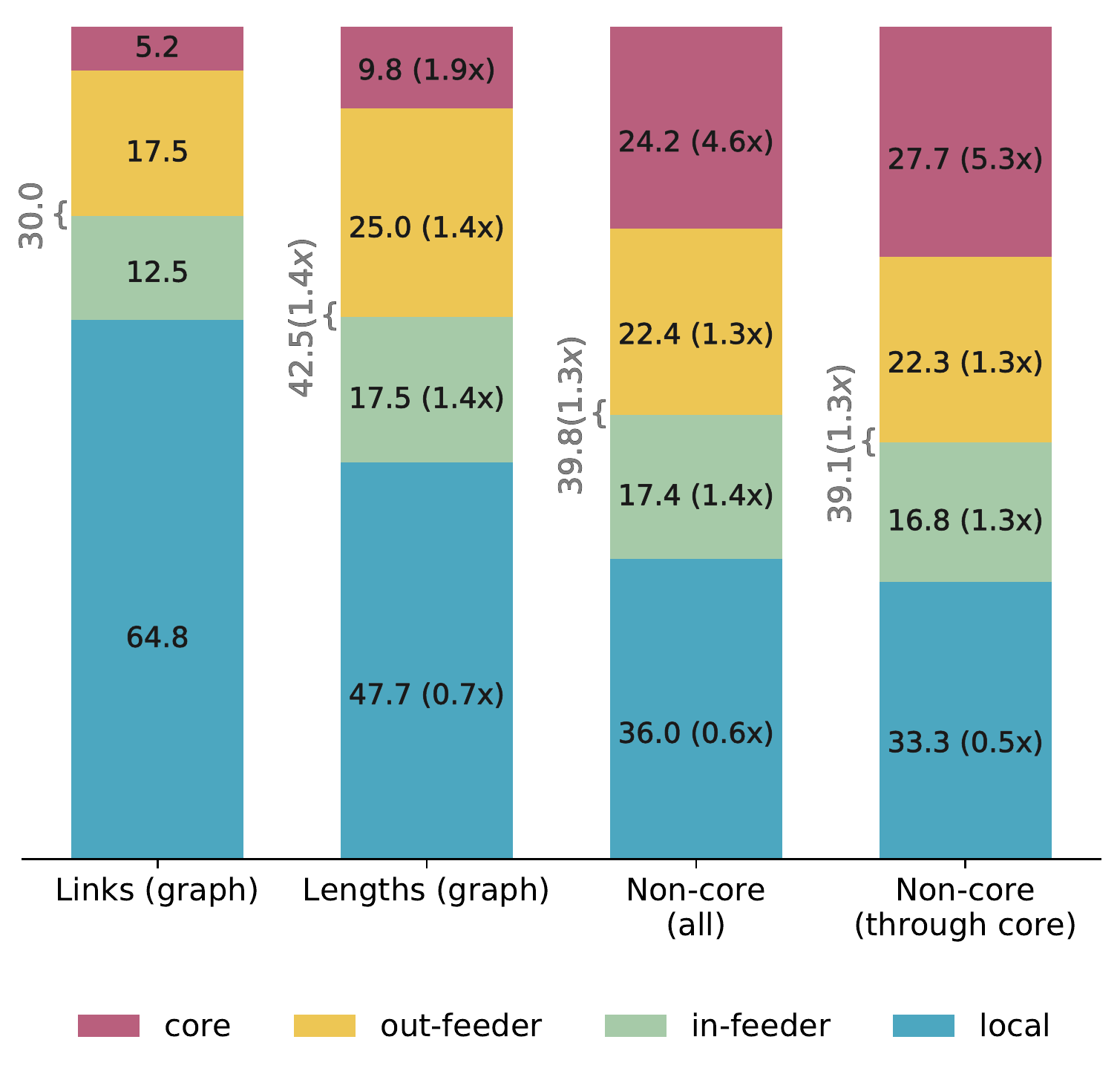}
		\caption{Percentage of links and shipping length by edge type in the original work (left) and with our updated minimum-route paths filtered with distance threshold $\alpha=1.15$ (right). The role of the core was simultaneously \emph{underestimated} for all shortest paths between non-core ports and \emph{overestimated} for only paths between non-core ports that pass through at least 1 core link. The first bar in each plot shows the percentage of links in the graph that fall into each category (sum of in and out feeders in gray to the left of each bar in the right plot). The second bar shows the percentage of total shipping length for links in each category in the graph. The numbers in parentheses are the length percentage divided by the link percentage. The majority of links in both representations are local (64.1\% and 64.8\%), but local links make up relatively less of the total length among all links in the graphs. The third bar in each plot shows the percentage of length by category for all paths between non-core ports (75\% of shortest paths in the previous analysis vs. 26\% here), while the fourth bar shows the same quantity but only for paths that pass through at least 1 core link (25\% previously vs. 74\% here). Using shortest paths through the undirected co-route graph we find that the role of the core was under-estimated for all paths (16.6\% vs. 24.2\%). However, the length to link percentage ratios are similar for all categories (core: 5.2 vs. 4.6; feeder: 1.7 vs. 1.3; local: 0.4 vs. 0.5). Limiting to only paths that pass through the structural core, the role of the core appears to have been overestimated (62.0\% vs. 27.7\%) and the role of local links was underestimated in the previous work (4.6\% vs. 33.3\%). In this case, the link to length percentage ratio for the core is much smaller (19.4 vs. 5.3), while the feeder (1.0 vs. 1.3) and local (0.1 vs. 0.5) ratios are again similar. Out-feeder links, which are links that leave the structural core, make up a higher percentage of both link and lengths than in-feeder links, which go from outside the core to inside.}
		\label{fig:linkandlengthpct}
	\end{figure}

	Figure~\ref{fig:linkandlengthpct} shows reproduced results from \cite{xu2020modular} (left) and results computed using minimum-route paths (right). The first two bars in each plot show the percent of links in the graph (the undirected co-route and path graphs, respectively) that fall into each category (leftmost bar) and the percent of total shipping length in each category (second bar from left). Both link percentages (3.2\% compared to 5.2\%) and length percentages (6.4\% compared to 9.8\%) in the path graph are slightly larger than those reported in \cite{xu2020modular}, suggesting that the role of core links in the graph structure was underestimated in the previous work. We also report that out-feeder links make up a larger percentage than in-feeder links in the path graph representation.
	
	The third and fourth bars in the left plot of Figure~\ref{fig:linkandlengthpct} represent length percentages for all shortest paths between pairs of non-core nodes (third bar) and only those paths that include at least one core link (rightmost bar). In the right plot we report the same quantities for minimum-route paths using distance threshold value $\alpha=1.15$ (more values of $\alpha$ reported in Appendix~\ref{app:core}). Only 25\% of shortest paths through the undirected co-route graph pass through the structural core. In contrast, 75\% of the filtered minimum-route paths pass through at least 1 core link. This difference between the paths helps explain the somewhat counterintuitive result that the role of core links was \emph{underestimated} for all paths between non-core ports (16.6\% vs. 24.2\%), but \emph{overestimated} for only the paths between non-core ports that pass through at least 1 core link (62.0\% vs. 27.7\%). In both cases the role of local links was underestimated; these links make up almost one third of the length in both sets of minimum-route paths between non-core ports.
	
	We also follow \cite{xu2020modular} by comparing the length-to-link percentage ratios for each of the length percentages (numbers in parenthesis in Figure~\ref{fig:linkandlengthpct}). Overall these ratios are similar between the two representations. The exception is the role of core links in mediating paths between non-core ports that pass through the core, where the previously reported length-to-link percentage ratio was 19.4, while the ratio in our analysis is 5.3.
	
	The difference in estimates of the role of core links can be explained in part by the choice of representation. When constructing the undirected co-route graph, ports that would require multiple intermediaries to reach one another based on the shipping routes are given undirected connections in the undirected co-route graph. Thus when shortest paths are computed, the number of intermediary nodes and links traversed is greatly reduced, since they are bypassed by direct connections created by the undirected co-route graph construction. In contrast, in the minimum-route paths these intermediary ports must be traversed in the order they appear in routes, meaning that local links are not avoided. This has two implications for our analysis. First, as stated above, local and feeder links are avoided in the shortest paths through the undirected co-route graph since direct edges are drawn between some pairs of ports that do not have any direct connections in the liner shipping service routes. Second, the size of the set of paths analyzed in the fourth bar changes from 25\% of the paths between non-core ports in the original work to 75\% in our analysis, which shows that core links are indispensable in supporting cargo transportation between non-core ports. Indeed, core links take up an even higher percentage of the total shipping length of paths between non-core ports that travel through the core (fourth bar, right plot) than in all paths between non-core ports (third bar, right plot), while the difference was overestimated in previous work (left plot). This helps explain the counterintuitive result that the previous analysis simultaneously overestimated and underestimated the role of the core links: using shortest paths through the undirected co-route representation made it so that the core was more easily avoided in the full set of paths between non-core ports, but also made it so that the set of paths that did pass through the core were biased towards using core links. This also explains why the third and fourth bars in the right plot are more similar to one another than those in the left plot.
	
	Despite the differences in the analyses, the general result from the previous analysis still holds. Our minimum-route path based analysis suggests that the structural core identified in the previous work does play an outsized role in mediating possible paths for cargo to take through the shipping network. However, we have found that quantifying the role of of this core using shortest paths through the undirected co-route graph representation simultaneously biases against and toward the core links depending on whether the paths being analyzed travel through the core at least once.
	
	\subsection{Route Betweenness}
	\label{sec:results-betw}
	
	\begin{figure}[!ht]
		
		\includegraphics[scale=0.45, center=-1.5cm]{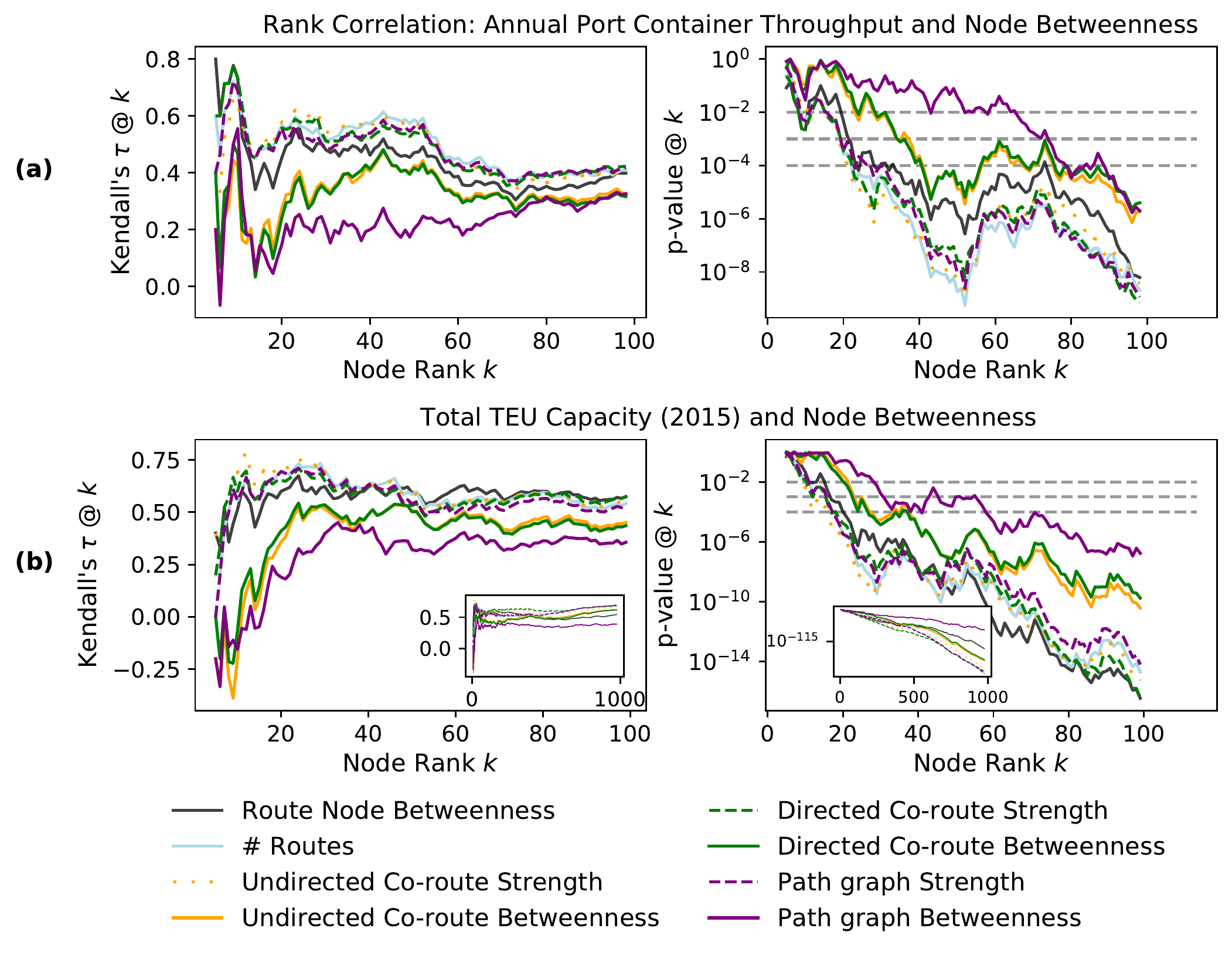}
		
		\rule{12cm}{0.05pt}
		
		\includegraphics[scale=0.45, center]{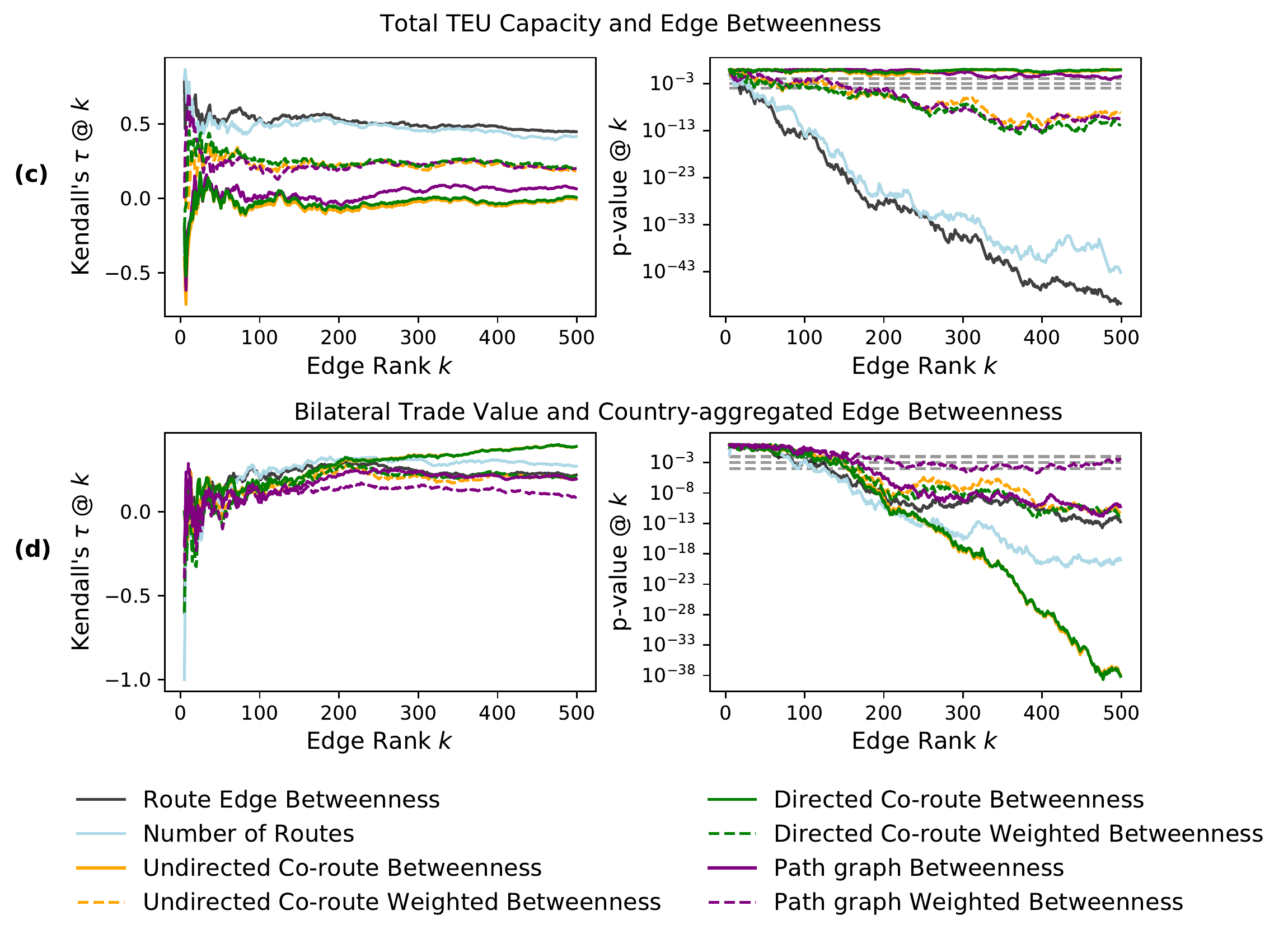}
		
		\caption{Rank correlations for port (a-b) and edge (c-d) importance measures. Ranking ports by number of routes they participates in has the strongest correlation with the top 100 ports by container throughput, and port rankings based on route betweenness correlate more strongly with ranking by container throughput than either topological betweenness measure. Ranking by route node betweenness correlates strongly with the top 100 ports by total TEU capacity. Correlation of edge rankings with route edge betweenness and total TEU capacity is strongly positive. Aggregating edge betweenness to the country level, betweenness in the directed and undirected co-route graphs correlates more strongly with bilateral trade than route edge betweenness or path graph betweenness. Dashed lines in the right plots indicate $p=0.01, 0.001$, and $0.0001$.}
		\label{fig:betweenness-kt}
	\end{figure}

	In this section we evaluate \emph{route betweenness} as described in Section~\ref{sec:methods-betw}, comparing our minimum-route path-based measures of route node and edge betweenness with topological centrality measures in the co-route and path graph representations. 
	
	We evaluate route node betweenness by measuring the correlation of the port rankings it produces with external data. We use as our baseline for comparison the top 100 ports based on container throughput downloaded from Lloyd's List Intelligence, a leading maritime shipping analyst service.\footnote{\url{https://www.lloydslistintelligence.com/}. Note that the top 100 container ports all together account for about 80\% of world's total container throughput each year. In fact, we use the top 98 ports because the ports at Ambarli and Dandong do not appear in the Alphaliner data. Further, the ports Keelung and Taipei are combined in the top 100 dataset, while they are separate in our shipping routes. Where applicable we use the minimum ranking between the two ports.} From this list we construct a rank vector $t=<1, 2, \cdots, 100>$ for the top container throughput ports, where the entry $t_i$ corresponds to the port $i$ with the $ith$ highest throughput. Then, we compute rankings of all ports based on route node betweenness centrality, (weighted) degree and (inverse weighted) betweenness centrality in both the path and undirected co-route graphs, and the count of the number of routes in which a node appears, where we define the weight of an edge to be the total number of times the edge appears across all of the routes. For each centrality ranking, we construct a rank vector $r$, where the entry $r_i$ is the ranking in the respective centrality measure for the port with the $i$th highest container throughput. The result is 8 rank vectors where each entry corresponds to the same port across all vectors. Finally, we compute Kendall's $\tau$ rank correlation \cite{kendall1970rank, 2020SciPy-NMeth} between the top container throughput ranking and each centrality ranking over a sliding window increasing in rank $k$.
	
	Figure~\ref{fig:betweenness-kt}a shows the results of this analysis. All 3 centrality measures correlate positively with container throughput, consistent with previous results \cite{xu2020modular}. The number of routes that a port appears in consistently has the largest rank correlation with the top 100 container throughput ports, and the strength (weighted degree) rankings are better correlated than any of the betweenness rankings. However, the correlation coefficient for the route node betweenness ranking is consistently larger than for the other betweenness measures for all values of k, and p-values on the coefficients are significant at $p=0.0001$ after the top 30. These results suggest that when measuring port importance, our route node betweenness measure is more consistent with a measure of importance external to the network than topological betweenness centralities, while simpler measures like weighted degree or the number of routes a port participates in are better correlates than centrality.
	
	We repeat this process again in Figure~\ref{fig:betweenness-kt}b, but this time using the total TEU capacity of all routes that a node participates in based on the data described in Section~\ref{sec:data}. The top 100 TEU Capacity port ranks are shown in the main plots, while the inset plots show correlation over all port ranks. Note that the strength rankings are determined by the total edges a node participates in over all routes, which does not include the TEU capacity information. Results are similar to the top 100 container throughput, where the strength and number of routes measures have consistently strong correlations, and route node betweenness correlates better than the other betweenness measures.
	
	We take a similar approach to evaluating route edge betweenness, computing the rank correlation between two external rankings and 7 edge centrality measures: route edge betweenness, (inverse weighted) edge betweenness in each of the directed and undirected coroute and path graphs. However, we must take care to properly evaluate the rankings given that edges in the directed co-route and path graph are directed, while edges in the undirected co-route graph are not. We achieve this by adding the values for the edge in both directions together, then orienting each edge across all of the rankings so the nodes are sorted alphabetically. For example, if the edges $(i,j)$ and $(j,i)$, $i<j$ both exist in one of the directed representations, we compute the sum of the measure of interest (e.g. betweenness) on both edges, then assign it to the single undirected edge $(i,j)$.
	
	The first external ranking is the sum of TEU capacity for all of the routes in which each edge appears, the same data as in Figure~\ref{fig:betweenness-kt}b. We construct a rank vector based on TEU capacity using this data. Then we construct rank vectors based on the edge centrality measures, and again compute Kendall's $\tau$ correlation between the capacity ranking and each of the centrality rankings. Results are shown in Figure~\ref{fig:betweenness-kt}c. Route edge betweenness is consistently the best correlated with edge TEU capacity. The inverse-weighted topological edge centralities correlate positively and reach low p-values by the top 500 edges, while the unweighted topological centralities hover around neutral and insignificant coefficients throughout.
	
	The second external ranking is the bilateral trade value between countries. This analysis is of practical relevance to understanding how the structural connectivity of the global liner shipping network is associated with international trade, given the fact that liner shipping accounts for about 70\% of global seaborne trade by value \cite{xu2020modular}. Since our edges are at the port level, we first aggregate the (now undirected) edge betweenness values by mapping each port to its country, then keeping a list of edge betweennness values for each pair of countries that have an edge. We then compute the rank correlation between the bilateral trade ranking and the centrality rankings, which we report in Firgure~\ref{fig:betweenness-kt}d. 
	
	In this case, the directed and undirected co-route graph edge betweenness rankings are most strongly correlated with the country level trade rankings. There is an intuitive reason for the co-route graph betweenness measures to be the strongest: the country-level rankings are based on bilateral trade without specific information about who mediates relationships between countries. When the routes are transformed into fully connected and undirected graphs in the undirected co-route graph, the bilateral relationships between the countries are maintained, but the more fine-grained information about who mediates trades---in terms of maritime transportation---between the countries is lost.
	
	\begin{figure}[!ht]
		\centering
		\includegraphics[scale=0.55]{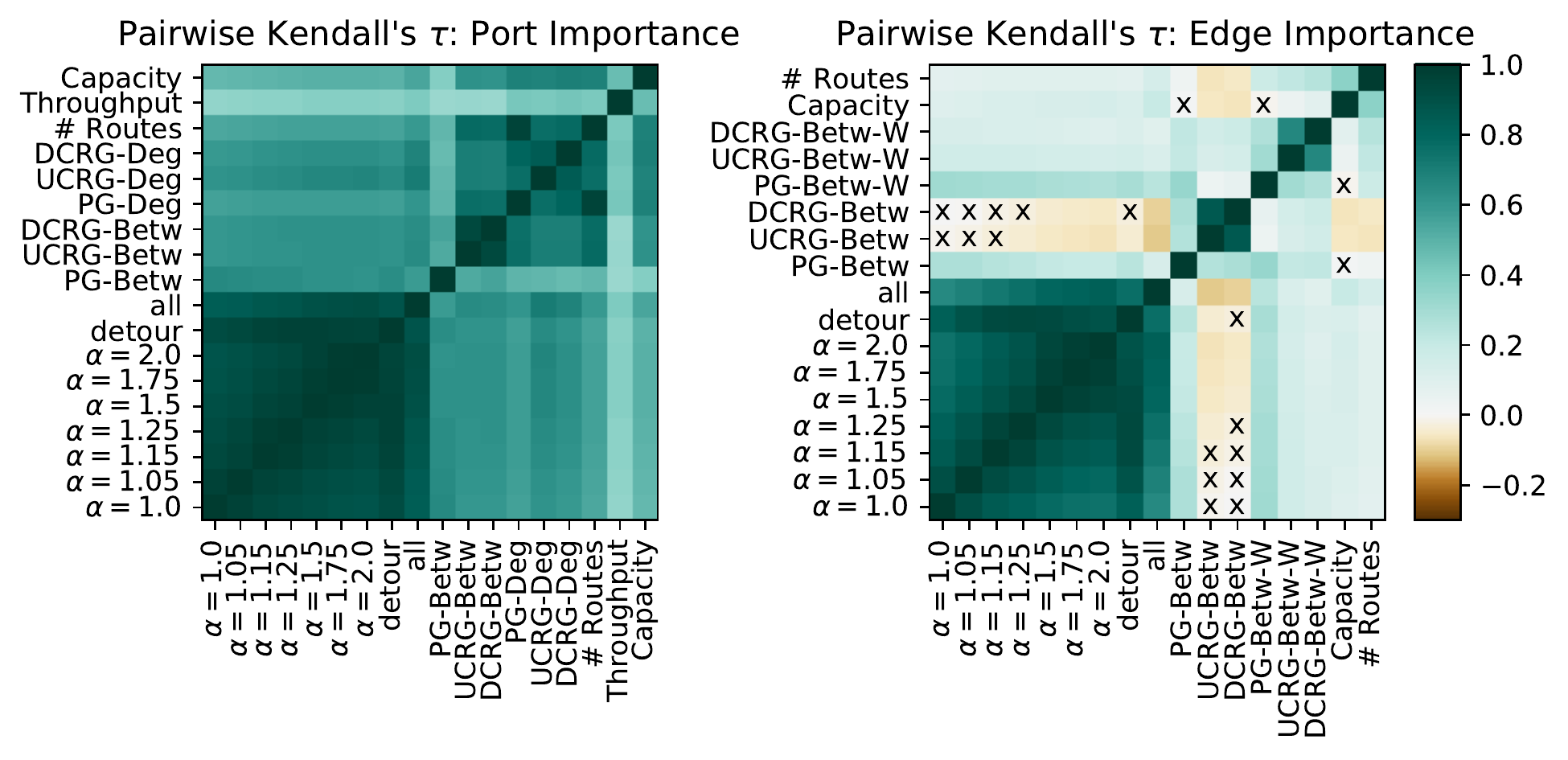}
		\caption{Pairwise Kendall's $\tau$ rank correlation for port rankings between all importance measures for nodes (left) and edges (right). "Capacity" means ranking by the sum of TEU capacity of all routes where a node or edge appears; "\# Routes" means ranking by the total number of routes in which the node or edge appears at least once; and "Throughput" means ranking the top 100 ports by container throughput. "DCRG" means directed co-route graph, "UCRG" means undirected co-route graph, and "betw" corresponds to betweenness and "deg" to degree. "all" refers to the full set of minimum-route paths without filtering and "detour" refers to detour factor filtering. Route node and edge betweenness measures correlate strongly across all values of shipping distance filtering threshold $\alpha$. All pairs of node importance measures have positive correlations. Edge betwenness in the directed and undirected co-route graphs are either neutral or slightly negatively correlated with other edge importance measures. Almost all p-values on the rank correlations coefficients were small; an \textbf{x} indicates that the p-value on the correlation coefficient was \emph{not} significant at $p=0.01$. Note that the correlations are symmetric.}
		\label{fig:pairwise-kt}
	\end{figure}
	
	Finally, we report the pairwise Kendall's $\tau$ for all importance measures in Figure~\ref{fig:pairwise-kt}. As expected, route node and edge betweenness for different values of distance filtering threshold $\alpha$ correlate highly with one another for both nodes and edges. This, along with the results in Figure~\ref{fig:allVSfiltered}, as well as Figure~\ref{fig:linkandlengthpct-app} in Appendix~\ref{app:core}, suggest that while the choice of $\alpha$ does change the route betweenness values and lower $\alpha$ reduces shipping distances, results appear to be robust to this parameter. All pairs of node importance measures have positive and significant (at least $p<0.01$) rank correlation coefficients. The edge betweenness rank correlations for the directed and undirected co-route graphs have neutral and even slightly negative correlations with the unweighted route edge betweenness measures, suggesting that these measures are indeed capturing different kinds of edge importance.
	
	Taken together, these results indicate that node and edge importance measures that take the shipping route data into account---including both our proposed route betweenness measures as well as simple counts of appearances in routes---correlate with external rankings as well as or better than measures that use shortest paths defined over the network structure. However, in some cases, such as when aggregating importance measures from ports up to countries, importance measures derived from the structure of the denser co-route graph representations may be better correlates than the route betweenness measures.
	
	\section{Conclusion}
	We presented analysis of liner shipping service route data using multiple network representations. We showed that the choice of representation has implications for the paths that can be inferred from the data, and that the choice of paths is important to analyzing the role of a structural core in the global maritime shipping network. Our analysis using an alternative set of paths, which we called minimum-route paths and compute using an algorithm called \method, suggests that previous work underestimated the role of core links in paths between non-core ports and overestimated the role of core links in the subset of paths that passed through at least one core link. Based on this analysis we also found that previous work underestimated the importance of local links. Despite this misestimation, the main conclusion from the previous work, that the structural core plays an outsized role in mediating navigation of cargo through the network, still follows from our analysis. Finally, we used our minimum-route paths to compute a measure of route betweenness centrality for both nodes and edges, and validated this measure against external measures of port and link importance, finding that our measure is at least as good as other indicators for throughput and capacity based node and link ranking, but simpler network indicators are better correlated with country-aggregated link importance.
	
	Our results suggest several criteria for choosing a representation when analyzing liner shipping service route data. If the research question is principally focused on dyadic trade relationships between entities, a coarser grained representation, such as the directed and undirected co-route graphs studied here, may be a reasonable and potentially advantageous representation. However, if the goal is to study the movement of cargo through the network, then either analyzing the routes themselves---as in route node or edge betweenness---or a representation that respects the directionality and direct connections in the network---as in the path graph---is likely to produce more accurate results.
	
	In future work, results should be compared with fine-grained ship and cargo movement data that was not the focus of this study. In particular, our path-based analysis, though an improvement over the undirected co-route graph analysis, does not take the timing of ship movements into account. It is well known that the temporal ordering of edge appearances can break apparent transitivity in network dynamics \cite{scholtes2017when, lambiotte2019networks}. Future analyses should also combine liner shipping service routes with data that captures the temporal patterns of ship movements, such as AIS data, to further our understanding of temporally viable minimum-route paths by ensuring paths are \emph{time-respecting} \cite{scholtes2016higherorder}. This could have important implications for which minimum-route paths through the network are truly viable in practice, since the temporal ordering of the trips could both limit the overall set of paths, as well as significantly alter the amount of time a path would take to realize.
	
	
	\begin{backmatter}
		
		\section*{Acknowledgements}
		TL acknowledges David Liu, Adina Gitomer, and Chia-Hung Yang for conversations about computing minimum-route paths; Brennan Klein and Ryan Gallagher for advice on interpreting and visualizing results; Harrison Hartle for comments on the draft manuscript; and Leo Torres for discussion about runtime analysis.   
		
		\section*{Funding}
		TL and TER are funded by in part by National Science Foundation grant IIS-1741197 and by the Combat Capabilities Development Command Army Research Laboratory through Cooperative Agreement W911NF-13-2-0045 and U.S. Army Research Lab Cyber Security CRA. The views and conclusions contained in this document are those of the authors and should not be interpreted as representing the official policies, either expressed or implied, of the Combat Capabilities Development Command Army Research Laboratory or the U.S. Government. The U.S. Government is authorized to reproduce and distribute reprints for Government purposes not withstanding any copyright notation hereon.
		
		MX is funded by the National Natural Science Foundation of China (Project Number: 72101046) and the Fundamental Research Funds for the Central Universities (DUT20RC(3)046) in China.
		
		\section*{Availability of data and materials}
		Raw data on world liner shipping services were provided by a third-party commercial database (Alphaliner, \url{https://www.alphaliner.com/}, one of the world’s leading databases in the liner shipping industry) and were used under the license for the current study, and so are not publicly available. Data on the nautical distance between ports are publicly available in: \url{https://www.searates.com/services/distances-time}. Data on countries’ international trade value and country pairs’ bilateral trade value are publicly available in: \url{https://comtrade.un.org/data}. Source data are provided with this paper.
		
		\section*{Competing interests}
		The authors declare that they have no competing interests.
		
		\section*{Authors' contributions}
		TL designed, implemented, and analyzed the IMR procedure; carried out all analyses; and wrote the first draft of the manuscript. MX contributed to writing and editing the manuscript; provided expertise on maritime shipping that informed all design and analysis throughout the paper; and provided all of the data, including the liner shipping service routes through an agreement with Alphaliner. TER contributed to writing and editing the paper and provided feedback on intermediate results and ideas throughout the project.
		
		
		\bibliographystyle{bmc-mathphys} 
		\bibliography{references}      


\begin{thebibliography}{30}
\ifx \bisbn   \undefined \def \bisbn  #1{ISBN #1}\fi
\ifx \binits  \undefined \def \binits#1{#1}\fi
\ifx \bauthor  \undefined \def \bauthor#1{#1}\fi
\ifx \batitle  \undefined \def \batitle#1{#1}\fi
\ifx \bjtitle  \undefined \def \bjtitle#1{#1}\fi
\ifx \bvolume  \undefined \def \bvolume#1{\textbf{#1}}\fi
\ifx \byear  \undefined \def \byear#1{#1}\fi
\ifx \bissue  \undefined \def \bissue#1{#1}\fi
\ifx \bfpage  \undefined \def \bfpage#1{#1}\fi
\ifx \blpage  \undefined \def \blpage #1{#1}\fi
\ifx \burl  \undefined \def \burl#1{\textsf{#1}}\fi
\ifx \doiurl  \undefined \def \doiurl#1{\textsf{#1}}\fi
\ifx \betal  \undefined \def \betal{\textit{et al.}}\fi
\ifx \binstitute  \undefined \def \binstitute#1{#1}\fi
\ifx \binstitutionaled  \undefined \def \binstitutionaled#1{#1}\fi
\ifx \bctitle  \undefined \def \bctitle#1{#1}\fi
\ifx \beditor  \undefined \def \beditor#1{#1}\fi
\ifx \bpublisher  \undefined \def \bpublisher#1{#1}\fi
\ifx \bbtitle  \undefined \def \bbtitle#1{#1}\fi
\ifx \bedition  \undefined \def \bedition#1{#1}\fi
\ifx \bseriesno  \undefined \def \bseriesno#1{#1}\fi
\ifx \blocation  \undefined \def \blocation#1{#1}\fi
\ifx \bsertitle  \undefined \def \bsertitle#1{#1}\fi
\ifx \bsnm \undefined \def \bsnm#1{#1}\fi
\ifx \bsuffix \undefined \def \bsuffix#1{#1}\fi
\ifx \bparticle \undefined \def \bparticle#1{#1}\fi
\ifx \barticle \undefined \def \barticle#1{#1}\fi
\ifx \bconfdate \undefined \def \bconfdate #1{#1}\fi
\ifx \botherref \undefined \def \botherref #1{#1}\fi
\ifx \url \undefined \def \url#1{\textsf{#1}}\fi
\ifx \bchapter \undefined \def \bchapter#1{#1}\fi
\ifx \bbook \undefined \def \bbook#1{#1}\fi
\ifx \bcomment \undefined \def \bcomment#1{#1}\fi
\ifx \oauthor \undefined \def \oauthor#1{#1}\fi
\ifx \citeauthoryear \undefined \def \citeauthoryear#1{#1}\fi
\ifx \endbibitem  \undefined \def \endbibitem {}\fi
\ifx \bconflocation  \undefined \def \bconflocation#1{#1}\fi
\ifx \arxivurl  \undefined \def \arxivurl#1{\textsf{#1}}\fi
\csname PreBibitemsHook\endcsname

\bibitem{hu2009empirical}
\begin{barticle}
\bauthor{\bsnm{Hu}, \binits{Y.}},
\bauthor{\bsnm{Zhu}, \binits{D.}}:
\batitle{Empirical analysis of the worldwide maritime transportation network}.
\bjtitle{Physica A: Statistical Mechanics and its Applications}
\bvolume{388}(\bissue{10}),
\bfpage{2061}--\blpage{2071}
(\byear{2009}).
doi:\doiurl{10.1016/j.physa.2008.12.016}
\end{barticle}
\endbibitem

\bibitem{kaluza2010complex}
\begin{barticle}
\bauthor{\bsnm{Kaluza}, \binits{P.}},
\bauthor{\bsnm{K{\"o}lzsch}, \binits{A.}},
\bauthor{\bsnm{Gastner}, \binits{M.T.}},
\bauthor{\bsnm{Blasius}, \binits{B.}}:
\batitle{The complex network of global cargo ship movements}.
\bjtitle{Journal of The Royal Society Interface}
\bvolume{7}(\bissue{48}),
\bfpage{1093}--\blpage{1103}
(\byear{2010}).
doi:\doiurl{10.1098/rsif.2009.0495}
\end{barticle}
\endbibitem

\bibitem{ducruet2010centrality}
\begin{barticle}
\bauthor{\bsnm{Ducruet}, \binits{C.}},
\bauthor{\bsnm{Lee}, \binits{S.-W.}},
\bauthor{\bsnm{Ng}, \binits{A.K.Y.}}:
\batitle{Centrality and vulnerability in liner shipping networks: Revisiting
  the {{Northeast Asian}} port hierarchy}.
\bjtitle{Maritime Policy \& Management}
\bvolume{37}(\bissue{1}),
\bfpage{17}--\blpage{36}
(\byear{2010}).
doi:\doiurl{10.1080/03088830903461175}
\end{barticle}
\endbibitem

\bibitem{ducruet2012maritime}
\begin{barticle}
\bauthor{\bsnm{Ducruet}, \binits{C.}},
\bauthor{\bsnm{Zaidi}, \binits{F.}}:
\batitle{Maritime constellations: A complex network approach to shipping and
  ports}.
\bjtitle{Maritime Policy \& Management}
\bvolume{39}(\bissue{2}),
\bfpage{151}--\blpage{168}
(\byear{2012}).
doi:\doiurl{10.1080/03088839.2011.650718}
\end{barticle}
\endbibitem

\bibitem{ducruet2012worldwide}
\begin{barticle}
\bauthor{\bsnm{Ducruet}, \binits{C.}},
\bauthor{\bsnm{Notteboom}, \binits{T.}}:
\batitle{The worldwide maritime network of container shipping: Spatial
  structure and regional dynamics}.
\bjtitle{Global Networks}
\bvolume{12}(\bissue{3}),
\bfpage{395}--\blpage{423}
(\byear{2012}).
doi:\doiurl{10.1111/j.1471-0374.2011.00355.x}
\end{barticle}
\endbibitem

\bibitem{ducruet2013network}
\begin{barticle}
\bauthor{\bsnm{Ducruet}, \binits{C.}}:
\batitle{Network diversity and maritime flows}.
\bjtitle{Journal of Transport Geography}
\bvolume{30},
\bfpage{77}--\blpage{88}
(\byear{2013}).
doi:\doiurl{10.1016/j.jtrangeo.2013.03.004}
\end{barticle}
\endbibitem

\bibitem{xu2014improving}
\begin{bchapter}
\bauthor{\bsnm{Xu}, \binits{J.}},
\bauthor{\bsnm{Wickramarathne}, \binits{T.L.}},
\bauthor{\bsnm{Chawla}, \binits{N.V.}},
\bauthor{\bsnm{Grey}, \binits{E.K.}},
\bauthor{\bsnm{Steinhaeuser}, \binits{K.}},
\bauthor{\bsnm{Keller}, \binits{R.P.}},
\bauthor{\bsnm{Drake}, \binits{J.M.}},
\bauthor{\bsnm{Lodge}, \binits{D.M.}}:
\bctitle{Improving management of aquatic invasions by integrating shipping
  network, ecological, and environmental data: Data mining for social good}.
In: \bbtitle{Proceedings of the 20th ACM SIGKDD International Conference on
  Knowledge Discovery and Data Mining},
pp. \bfpage{1699}--\blpage{1708}
(\byear{2014}).
doi:\doiurl{10.1145/2623330.2623364}
\end{bchapter}
\endbibitem

\bibitem{li2015centrality}
\begin{barticle}
\bauthor{\bsnm{Li}, \binits{Z.}},
\bauthor{\bsnm{Xu}, \binits{M.}},
\bauthor{\bsnm{Shi}, \binits{Y.}}:
\batitle{Centrality in global shipping network basing on worldwide shipping
  areas}.
\bjtitle{GeoJournal}
\bvolume{80}(\bissue{1}),
\bfpage{47}--\blpage{60}
(\byear{2015}).
doi:\doiurl{10.1007/s10708-014-9524-3}
\end{barticle}
\endbibitem

\bibitem{xu2015evolution}
\begin{barticle}
\bauthor{\bsnm{Xu}, \binits{M.}},
\bauthor{\bsnm{Li}, \binits{Z.}},
\bauthor{\bsnm{Shi}, \binits{Y.}},
\bauthor{\bsnm{Zhang}, \binits{X.}},
\bauthor{\bsnm{Jiang}, \binits{S.}}:
\batitle{Evolution of regional inequality in the global shipping network}.
\bjtitle{Journal of Transport Geography}
\bvolume{44},
\bfpage{1}--\blpage{12}
(\byear{2015}).
doi:\doiurl{10.1016/j.jtrangeo.2015.02.003}
\end{barticle}
\endbibitem

\bibitem{kojaku2019multiscale}
\begin{barticle}
\bauthor{\bsnm{Kojaku}, \binits{S.}},
\bauthor{\bsnm{Xu}, \binits{M.}},
\bauthor{\bsnm{Xia}, \binits{H.}},
\bauthor{\bsnm{Masuda}, \binits{N.}}:
\batitle{Multiscale core-periphery structure in a global liner shipping
  network}.
\bjtitle{Scientific Reports}
\bvolume{9}(\bissue{1}),
\bfpage{404}
(\byear{2019}).
doi:\doiurl{10.1038/s41598-018-35922-2}
\end{barticle}
\endbibitem

\bibitem{xu2020modular}
\begin{barticle}
\bauthor{\bsnm{Xu}, \binits{M.}},
\bauthor{\bsnm{Pan}, \binits{Q.}},
\bauthor{\bsnm{Muscoloni}, \binits{A.}},
\bauthor{\bsnm{Xia}, \binits{H.}},
\bauthor{\bsnm{Cannistraci}, \binits{C.V.}}:
\batitle{Modular gateway-ness connectivity and structural core organization in
  maritime network science}.
\bjtitle{Nature Communications}
\bvolume{11}(\bissue{1}),
\bfpage{2849}
(\byear{2020}).
doi:\doiurl{10.1038/s41467-020-16619-5}
\end{barticle}
\endbibitem

\bibitem{torres2021why}
\begin{barticle}
\bauthor{\bsnm{Torres}, \binits{L.}},
\bauthor{\bsnm{Blevins}, \binits{A.S.}},
\bauthor{\bsnm{Bassett}, \binits{D.}},
\bauthor{\bsnm{{Eliassi-Rad}}, \binits{T.}}:
\batitle{The {{Why}}, {{How}}, and {{When}} of {{Representations}} for
  {{Complex Systems}}}.
\bjtitle{SIAM Rev.}
\bvolume{63}(\bissue{3}),
\bfpage{435}--\blpage{485}
(\byear{2021}).
doi:\doiurl{10.1137/20M1355896}
\end{barticle}
\endbibitem

\bibitem{chodrow2020configuration}
\begin{barticle}
\bauthor{\bsnm{Chodrow}, \binits{P.S.}}:
\batitle{Configuration models of random hypergraphs}.
\bjtitle{Journal of Complex Networks}
\bvolume{8}(\bissue{3}),
\bfpage{018}
(\byear{2020}).
doi:\doiurl{10.1093/comnet/cnaa018}
\end{barticle}
\endbibitem

\bibitem{battiston2020networksa}
\begin{barticle}
\bauthor{\bsnm{Battiston}, \binits{F.}},
\bauthor{\bsnm{Cencetti}, \binits{G.}},
\bauthor{\bsnm{Iacopini}, \binits{I.}},
\bauthor{\bsnm{Latora}, \binits{V.}},
\bauthor{\bsnm{Lucas}, \binits{M.}},
\bauthor{\bsnm{Patania}, \binits{A.}},
\bauthor{\bsnm{Young}, \binits{J.-G.}},
\bauthor{\bsnm{Petri}, \binits{G.}}:
\batitle{Networks beyond pairwise interactions: Structure and dynamics}.
\bjtitle{Physics Reports}
\bvolume{874},
\bfpage{1}--\blpage{92}
(\byear{2020}).
doi:\doiurl{10.1016/j.physrep.2020.05.004}.
\arxivurl{2006.01764}
\end{barticle}
\endbibitem

\bibitem{scholtes2017when}
\begin{bchapter}
\bauthor{\bsnm{Scholtes}, \binits{I.}}:
\bctitle{When is a network a network?: Multi-order graphical model selection in
  pathways and temporal networks}.
In: \bbtitle{Proceedings of the 23rd ACM SIGKDD International Conference on
  Knowledge Discovery and Data Mining},
pp. \bfpage{1037}--\blpage{1046}
(\byear{2017}).
doi:\doiurl{10.1145/3097983.3098145}
\end{bchapter}
\endbibitem

\bibitem{lambiotte2019networks}
\begin{barticle}
\bauthor{\bsnm{Lambiotte}, \binits{R.}},
\bauthor{\bsnm{Rosvall}, \binits{M.}},
\bauthor{\bsnm{Scholtes}, \binits{I.}}:
\batitle{From networks to optimal higher-order models of complex systems}.
\bjtitle{Nature Physics}
\bvolume{15}(\bissue{4}),
\bfpage{313}--\blpage{320}
(\byear{2019}).
doi:\doiurl{10.1038/s41567-019-0459-y}
\end{barticle}
\endbibitem

\bibitem{saebi2020network}
\begin{barticle}
\bauthor{\bsnm{Saebi}, \binits{M.}},
\bauthor{\bsnm{Xu}, \binits{J.}},
\bauthor{\bsnm{Curasi}, \binits{S.R.}},
\bauthor{\bsnm{Grey}, \binits{E.K.}},
\bauthor{\bsnm{Chawla}, \binits{N.V.}},
\bauthor{\bsnm{Lodge}, \binits{D.M.}}:
\batitle{Network analysis of ballast-mediated species transfer reveals
  important introduction and dispersal patterns in the {{Arctic}}}.
\bjtitle{Scientific Reports}
\bvolume{10}(\bissue{1}),
\bfpage{19558}
(\byear{2020}).
doi:\doiurl{10.1038/s41598-020-76602-4}
\end{barticle}
\endbibitem

\bibitem{xu2016representing}
\begin{barticle}
\bauthor{\bsnm{Xu}, \binits{J.}},
\bauthor{\bsnm{Wickramarathne}, \binits{T.L.}},
\bauthor{\bsnm{Chawla}, \binits{N.V.}}:
\batitle{Representing higher-order dependencies in networks}.
\bjtitle{Science Advances}
\bvolume{2}(\bissue{5}),
\bfpage{1600028}--\blpage{1600028}
(\byear{2016}).
doi:\doiurl{10.1126/sciadv.1600028}
\end{barticle}
\endbibitem

\bibitem{barrett2008engineering}
\begin{bchapter}
\bauthor{\bsnm{Barrett}, \binits{C.}},
\bauthor{\bsnm{Bisset}, \binits{K.}},
\bauthor{\bsnm{Holzer}, \binits{M.}},
\bauthor{\bsnm{Konjevod}, \binits{G.}},
\bauthor{\bsnm{Marathe}, \binits{M.}},
\bauthor{\bsnm{Wagner}, \binits{D.}}:
\bctitle{Engineering {{Label}}-{{Constrained Shortest}}-{{Path Algorithms}}}.
In: \beditor{\bsnm{Fleischer}, \binits{R.}},
\beditor{\bsnm{Xu}, \binits{J.}} (eds.)
\bbtitle{Algorithmic {{Aspects}} in {{Information}} and {{Management}}}
vol. \bseriesno{5034},
pp. \bfpage{27}--\blpage{37}
(\byear{2008}).
doi:\doiurl{10.1007/978-3-540-68880-8\_5}
\end{bchapter}
\endbibitem

\bibitem{bast2010fast}
\begin{bchapter}
\bauthor{\bsnm{Bast}, \binits{H.}},
\bauthor{\bsnm{Carlsson}, \binits{E.}},
\bauthor{\bsnm{Eigenwillig}, \binits{A.}},
\bauthor{\bsnm{Geisberger}, \binits{R.}},
\bauthor{\bsnm{Harrelson}, \binits{C.}},
\bauthor{\bsnm{Raychev}, \binits{V.}},
\bauthor{\bsnm{Viger}, \binits{F.}}:
\bctitle{Fast {{Routing}} in {{Very Large Public Transportation Networks Using
  Transfer Patterns}}}.
In: \beditor{\bsnm{Hutchison}, \binits{D.}},
\beditor{\bsnm{Kanade}, \binits{T.}},
\beditor{\bsnm{Kittler}, \binits{J.}},
\beditor{\bsnm{Kleinberg}, \binits{J.M.}},
\beditor{\bsnm{Mattern}, \binits{F.}},
\beditor{\bsnm{Mitchell}, \binits{J.C.}},
\beditor{\bsnm{Naor}, \binits{M.}},
\beditor{\bsnm{Nierstrasz}, \binits{O.}},
\beditor{\bsnm{Pandu~Rangan}, \binits{C.}},
\beditor{\bsnm{Steffen}, \binits{B.}},
\beditor{\bsnm{Sudan}, \binits{M.}},
\beditor{\bsnm{Terzopoulos}, \binits{D.}},
\beditor{\bsnm{Tygar}, \binits{D.}},
\beditor{\bsnm{Vardi}, \binits{M.Y.}},
\beditor{\bsnm{Weikum}, \binits{G.}},
\beditor{\bsnm{{de Berg}}, \binits{M.}},
\beditor{\bsnm{Meyer}, \binits{U.}} (eds.)
\bbtitle{Algorithms \textendash{} {{ESA}} 2010}
vol. \bseriesno{6346},
pp. \bfpage{290}--\blpage{301}
(\byear{2010}).
doi:\doiurl{10.1007/978-3-642-15775-2\_25}
\end{bchapter}
\endbibitem

\bibitem{lozano2001shortest}
\begin{barticle}
\bauthor{\bsnm{Lozano}, \binits{A.}},
\bauthor{\bsnm{Storchi}, \binits{G.}}:
\batitle{Shortest viable path algorithm in multimodal networks}.
\bjtitle{Transportation Research Part A: Policy and Practice}
\bvolume{35}(\bissue{3}),
\bfpage{225}--\blpage{241}
(\byear{2001}).
doi:\doiurl{10.1016/S0965-8564(99)00056-}
\end{barticle}
\endbibitem

\bibitem{lewis2020algorithms}
\begin{barticle}
\bauthor{\bsnm{Lewis}, \binits{R.}}:
\batitle{Algorithms for {{Finding Shortest Paths}} in {{Networks}} with
  {{Vertex Transfer Penalties}}}.
\bjtitle{Algorithms}
\bvolume{13}(\bissue{11}),
\bfpage{269}
(\byear{2020}).
doi:\doiurl{10.3390/a13110269}
\end{barticle}
\endbibitem

\bibitem{ferone2019kcolor}
\begin{bchapter}
\bauthor{\bsnm{Ferone}, \binits{D.}},
\bauthor{\bsnm{Festa}, \binits{P.}},
\bauthor{\bsnm{Pastore}, \binits{T.}}:
\bctitle{The k-color shortest path problem}.
In: \beditor{\bsnm{Paolucci}, \binits{M.}},
\beditor{\bsnm{Sciomachen}, \binits{A.}},
\beditor{\bsnm{Uberti}, \binits{P.}} (eds.)
\bbtitle{Advances in Optimization and Decision Science for Society, Services
  and Enterprises}
vol. \bseriesno{3},
pp. \bfpage{367}--\blpage{376}.
\bpublisher{{Springer International Publishing}},
\blocation{{Cham}}
(\byear{2019}).
doi:\doiurl{10.1007/978-3-030-34960-8\_32}
\end{bchapter}
\endbibitem

\bibitem{bohmova2018computing}
\begin{barticle}
\bauthor{\bsnm{B{\"o}hmov{\'a}}, \binits{K.}},
\bauthor{\bsnm{H{\"a}fliger}, \binits{L.}},
\bauthor{\bsnm{Mihal{\'a}k}, \binits{M.}},
\bauthor{\bsnm{Pr{\"o}ger}, \binits{T.}},
\bauthor{\bsnm{Sacomoto}, \binits{G.}},
\bauthor{\bsnm{Sagot}, \binits{M.-F.}}:
\batitle{Computing and {{Listing}} st-{{Paths}} in {{Public Transportation
  Networks}}}.
\bjtitle{Theory of Computing Systems}
\bvolume{62}(\bissue{3}),
\bfpage{600}--\blpage{621}
(\byear{2018}).
doi:\doiurl{10.1007/s00224-016-9747-4}
\end{barticle}
\endbibitem

\bibitem{larock2020hypa}
\begin{bchapter}
\bauthor{\bsnm{LaRock}, \binits{T.}},
\bauthor{\bsnm{Nanumyan}, \binits{V.}},
\bauthor{\bsnm{Scholtes}, \binits{I.}},
\bauthor{\bsnm{Casiraghi}, \binits{G.}},
\bauthor{\bsnm{{Eliassi-Rad}}, \binits{T.}},
\bauthor{\bsnm{Schweitzer}, \binits{F.}}:
\bctitle{Hypa: Efficient detection of path anomalies in time series data on
  networks}.
In: \bbtitle{Proceedings of the 2020 SIAM International Conference on Data
  Mining},
pp. \bfpage{460}--\blpage{468}
(\byear{2020}).
doi:\doiurl{10.1137/1.9781611976236.52}
\end{bchapter}
\endbibitem

\bibitem{yang2018auniversal}
\begin{barticle}
\bauthor{\bsnm{Yang}, \binits{H.}},
\bauthor{\bsnm{Ke}, \binits{J.}},
\bauthor{\bsnm{Ye}, \binits{J.}}:
\batitle{A universal distribution law of network detour ratios}.
\bjtitle{Transportation Research Part C: Emerging Technologies}
\bvolume{96},
\bfpage{22}--\blpage{37}
(\byear{2018}).
doi:\doiurl{10.1016/j.trc.2018.09.012}
\end{barticle}
\endbibitem

\bibitem{blondel2008fast}
\begin{barticle}
\bauthor{\bsnm{Blondel}, \binits{V.D.}},
\bauthor{\bsnm{Guillaume}, \binits{J.-L.}},
\bauthor{\bsnm{Lambiotte}, \binits{R.}},
\bauthor{\bsnm{Lefebvre}, \binits{E.}}:
\batitle{Fast unfolding of communities in large networks}.
\bjtitle{Journal of Statistical Mechanics: Theory and Experiment}
\bvolume{2008}(\bissue{10}),
\bfpage{10008}
(\byear{2008}).
doi:\doiurl{10.1088/1742-5468/2008/10/P10008}
\end{barticle}
\endbibitem

\bibitem{kendall1970rank}
\begin{bbook}
\bauthor{\bsnm{Kendall}, \binits{M.G.}}:
\bbtitle{Rank Correlation Methods},
\bedition{4th ed} edn.
\bpublisher{{Griffin}},
\blocation{{London}}
(\byear{1970})
\end{bbook}
\endbibitem

\bibitem{2020SciPy-NMeth}
\begin{barticle}
\bauthor{\bsnm{Virtanen}, \binits{P.}},
\bauthor{\bsnm{Gommers}, \binits{R.}},
\bauthor{\bsnm{Oliphant}, \binits{T.E.}},
\bauthor{\bsnm{Haberland}, \binits{M.}},
\bauthor{\bsnm{Reddy}, \binits{T.}},
\bauthor{\bsnm{Cournapeau}, \binits{D.}},
\bauthor{\bsnm{Burovski}, \binits{E.}},
\bauthor{\bsnm{Peterson}, \binits{P.}},
\bauthor{\bsnm{Weckesser}, \binits{W.}},
\bauthor{\bsnm{Bright}, \binits{J.}},
\bauthor{\bsnm{{van der Walt}}, \binits{S.J.}},
\bauthor{\bsnm{Brett}, \binits{M.}},
\bauthor{\bsnm{Wilson}, \binits{J.}},
\bauthor{\bsnm{Millman}, \binits{K.J.}},
\bauthor{\bsnm{Mayorov}, \binits{N.}},
\bauthor{\bsnm{Nelson}, \binits{A.R.J.}},
\bauthor{\bsnm{Jones}, \binits{E.}},
\bauthor{\bsnm{Kern}, \binits{R.}},
\bauthor{\bsnm{Larson}, \binits{E.}},
\bauthor{\bsnm{Carey}, \binits{C.J.}},
\bauthor{\bsnm{Polat}, \binits{{\. I}.}},
\bauthor{\bsnm{Feng}, \binits{Y.}},
\bauthor{\bsnm{Moore}, \binits{E.W.}},
\bauthor{\bsnm{{VanderPlas}}, \binits{J.}},
\bauthor{\bsnm{Laxalde}, \binits{D.}},
\bauthor{\bsnm{Perktold}, \binits{J.}},
\bauthor{\bsnm{Cimrman}, \binits{R.}},
\bauthor{\bsnm{Henriksen}, \binits{I.}},
\bauthor{\bsnm{Quintero}, \binits{E.A.}},
\bauthor{\bsnm{Harris}, \binits{C.R.}},
\bauthor{\bsnm{Archibald}, \binits{A.M.}},
\bauthor{\bsnm{Ribeiro}, \binits{A.H.}},
\bauthor{\bsnm{Pedregosa}, \binits{F.}},
\bauthor{\bsnm{{van Mulbregt}}, \binits{P.}},
\bauthor{\bsnm{{SciPy 1.0 Contributors}}}:
\batitle{{{SciPy} 1.0: Fundamental Algorithms for Scientific Computing in
  Python}}.
\bjtitle{Nature Methods}
\bvolume{17},
\bfpage{261}--\blpage{272}
(\byear{2020}).
doi:\doiurl{10.1038/s41592-019-0686-2}
\end{barticle}
\endbibitem

\bibitem{scholtes2016higherorder}
\begin{botherref}
\oauthor{\bsnm{Scholtes}, \binits{I.}},
\oauthor{\bsnm{Wider}, \binits{N.}},
\oauthor{\bsnm{Garas}, \binits{A.}}:
Higher-order aggregate networks in the analysis of temporal networks: Path
  structures and centralities.
The European Physical Journal B
\textbf{89}(3)
(2016).
doi:\doiurl{10.1140/epjb/e2016-60663-0}
\end{botherref}
\endbibitem

\end{thebibliography}

\newcommand{\BMCxmlcomment}[1]{}

\BMCxmlcomment{

<refgrp>

<bibl id="B1">
  <title><p>Empirical Analysis of the Worldwide Maritime Transportation
  Network</p></title>
  <aug>
    <au><snm>Hu</snm><fnm>Y</fnm></au>
    <au><snm>Zhu</snm><fnm>D</fnm></au>
  </aug>
  <source>Physica A: Statistical Mechanics and its Applications</source>
  <pubdate>2009</pubdate>
  <volume>388</volume>
  <issue>10</issue>
  <fpage>2061</fpage>
  <lpage>-2071</lpage>
</bibl>

<bibl id="B2">
  <title><p>The Complex Network of Global Cargo Ship Movements</p></title>
  <aug>
    <au><snm>Kaluza</snm><fnm>P</fnm></au>
    <au><snm>K{\"o}lzsch</snm><fnm>A</fnm></au>
    <au><snm>Gastner</snm><fnm>MT</fnm></au>
    <au><snm>Blasius</snm><fnm>B</fnm></au>
  </aug>
  <source>Journal of The Royal Society Interface</source>
  <pubdate>2010</pubdate>
  <volume>7</volume>
  <issue>48</issue>
  <fpage>1093</fpage>
  <lpage>-1103</lpage>
</bibl>

<bibl id="B3">
  <title><p>Centrality and Vulnerability in Liner Shipping Networks: Revisiting
  the {{Northeast Asian}} Port Hierarchy</p></title>
  <aug>
    <au><snm>Ducruet</snm><fnm>C</fnm></au>
    <au><snm>Lee</snm><fnm>SW</fnm></au>
    <au><snm>Ng</snm><fnm>AK</fnm></au>
  </aug>
  <source>Maritime Policy \& Management</source>
  <pubdate>2010</pubdate>
  <volume>37</volume>
  <issue>1</issue>
  <fpage>17</fpage>
  <lpage>-36</lpage>
</bibl>

<bibl id="B4">
  <title><p>Maritime Constellations: A Complex Network Approach to Shipping and
  Ports</p></title>
  <aug>
    <au><snm>Ducruet</snm><fnm>C</fnm></au>
    <au><snm>Zaidi</snm><fnm>F</fnm></au>
  </aug>
  <source>Maritime Policy \& Management</source>
  <pubdate>2012</pubdate>
  <volume>39</volume>
  <issue>2</issue>
  <fpage>151</fpage>
  <lpage>-168</lpage>
</bibl>

<bibl id="B5">
  <title><p>The Worldwide Maritime Network of Container Shipping: Spatial
  Structure and Regional Dynamics</p></title>
  <aug>
    <au><snm>Ducruet</snm><fnm>C</fnm></au>
    <au><snm>Notteboom</snm><fnm>T</fnm></au>
  </aug>
  <source>Global Networks</source>
  <pubdate>2012</pubdate>
  <volume>12</volume>
  <issue>3</issue>
  <fpage>395</fpage>
  <lpage>-423</lpage>
</bibl>

<bibl id="B6">
  <title><p>Network Diversity and Maritime Flows</p></title>
  <aug>
    <au><snm>Ducruet</snm><fnm>C</fnm></au>
  </aug>
  <source>Journal of Transport Geography</source>
  <pubdate>2013</pubdate>
  <volume>30</volume>
  <fpage>77</fpage>
  <lpage>-88</lpage>
</bibl>

<bibl id="B7">
  <title><p>Improving Management of Aquatic Invasions by Integrating Shipping
  Network, Ecological, and Environmental Data: Data Mining for Social
  Good</p></title>
  <aug>
    <au><snm>Xu</snm><fnm>J</fnm></au>
    <au><snm>Wickramarathne</snm><fnm>TL</fnm></au>
    <au><snm>Chawla</snm><fnm>NV</fnm></au>
    <au><snm>Grey</snm><fnm>EK</fnm></au>
    <au><snm>Steinhaeuser</snm><fnm>K</fnm></au>
    <au><snm>Keller</snm><fnm>RP</fnm></au>
    <au><snm>Drake</snm><fnm>JM</fnm></au>
    <au><snm>Lodge</snm><fnm>DM</fnm></au>
  </aug>
  <source>Proceedings of the 20th ACM SIGKDD International Conference on
  Knowledge Discovery and Data Mining</source>
  <pubdate>2014</pubdate>
  <fpage>1699</fpage>
  <lpage>-1708</lpage>
</bibl>

<bibl id="B8">
  <title><p>Centrality in Global Shipping Network Basing on Worldwide Shipping
  Areas</p></title>
  <aug>
    <au><snm>Li</snm><fnm>Z</fnm></au>
    <au><snm>Xu</snm><fnm>M</fnm></au>
    <au><snm>Shi</snm><fnm>Y</fnm></au>
  </aug>
  <source>GeoJournal</source>
  <pubdate>2015</pubdate>
  <volume>80</volume>
  <issue>1</issue>
  <fpage>47</fpage>
  <lpage>-60</lpage>
</bibl>

<bibl id="B9">
  <title><p>Evolution of Regional Inequality in the Global Shipping
  Network</p></title>
  <aug>
    <au><snm>Xu</snm><fnm>M</fnm></au>
    <au><snm>Li</snm><fnm>Z</fnm></au>
    <au><snm>Shi</snm><fnm>Y</fnm></au>
    <au><snm>Zhang</snm><fnm>X</fnm></au>
    <au><snm>Jiang</snm><fnm>S</fnm></au>
  </aug>
  <source>Journal of Transport Geography</source>
  <pubdate>2015</pubdate>
  <volume>44</volume>
  <fpage>1</fpage>
  <lpage>-12</lpage>
</bibl>

<bibl id="B10">
  <title><p>Multiscale Core-Periphery Structure in a Global Liner Shipping
  Network</p></title>
  <aug>
    <au><snm>Kojaku</snm><fnm>S</fnm></au>
    <au><snm>Xu</snm><fnm>M</fnm></au>
    <au><snm>Xia</snm><fnm>H</fnm></au>
    <au><snm>Masuda</snm><fnm>N</fnm></au>
  </aug>
  <source>Scientific Reports</source>
  <pubdate>2019</pubdate>
  <volume>9</volume>
  <issue>1</issue>
  <fpage>404</fpage>
</bibl>

<bibl id="B11">
  <title><p>Modular Gateway-Ness Connectivity and Structural Core Organization
  in Maritime Network Science</p></title>
  <aug>
    <au><snm>Xu</snm><fnm>M</fnm></au>
    <au><snm>Pan</snm><fnm>Q</fnm></au>
    <au><snm>Muscoloni</snm><fnm>A</fnm></au>
    <au><snm>Xia</snm><fnm>H</fnm></au>
    <au><snm>Cannistraci</snm><fnm>CV</fnm></au>
  </aug>
  <source>Nature Communications</source>
  <pubdate>2020</pubdate>
  <volume>11</volume>
  <issue>1</issue>
  <fpage>2849</fpage>
</bibl>

<bibl id="B12">
  <title><p>The {{Why}}, {{How}}, and {{When}} of {{Representations}} for
  {{Complex Systems}}</p></title>
  <aug>
    <au><snm>Torres</snm><fnm>L</fnm></au>
    <au><snm>Blevins</snm><fnm>AS</fnm></au>
    <au><snm>Bassett</snm><fnm>D</fnm></au>
    <au><snm>{Eliassi-Rad}</snm><fnm>T</fnm></au>
  </aug>
  <source>SIAM Rev.</source>
  <pubdate>2021</pubdate>
  <volume>63</volume>
  <issue>3</issue>
  <fpage>435</fpage>
  <lpage>-485</lpage>
</bibl>

<bibl id="B13">
  <title><p>Configuration Models of Random Hypergraphs</p></title>
  <aug>
    <au><snm>Chodrow</snm><fnm>PS</fnm></au>
  </aug>
  <source>Journal of Complex Networks</source>
  <editor>Gleeson, James</editor>
  <pubdate>2020</pubdate>
  <volume>8</volume>
  <issue>3</issue>
  <fpage>cnaa018</fpage>
</bibl>

<bibl id="B14">
  <title><p>Networks beyond Pairwise Interactions: Structure and
  Dynamics</p></title>
  <aug>
    <au><snm>Battiston</snm><fnm>F</fnm></au>
    <au><snm>Cencetti</snm><fnm>G</fnm></au>
    <au><snm>Iacopini</snm><fnm>I</fnm></au>
    <au><snm>Latora</snm><fnm>V</fnm></au>
    <au><snm>Lucas</snm><fnm>M</fnm></au>
    <au><snm>Patania</snm><fnm>A</fnm></au>
    <au><snm>Young</snm><fnm>JG</fnm></au>
    <au><snm>Petri</snm><fnm>G</fnm></au>
  </aug>
  <source>Physics Reports</source>
  <pubdate>2020</pubdate>
  <volume>874</volume>
  <fpage>1</fpage>
  <lpage>-92</lpage>
</bibl>

<bibl id="B15">
  <title><p>When Is a Network a Network?: Multi-Order Graphical Model Selection
  in Pathways and Temporal Networks</p></title>
  <aug>
    <au><snm>Scholtes</snm><fnm>I</fnm></au>
  </aug>
  <source>Proceedings of the 23rd ACM SIGKDD International Conference on
  Knowledge Discovery and Data Mining</source>
  <pubdate>2017</pubdate>
  <fpage>1037</fpage>
  <lpage>-1046</lpage>
</bibl>

<bibl id="B16">
  <title><p>From Networks to Optimal Higher-Order Models of Complex
  Systems</p></title>
  <aug>
    <au><snm>Lambiotte</snm><fnm>R</fnm></au>
    <au><snm>Rosvall</snm><fnm>M</fnm></au>
    <au><snm>Scholtes</snm><fnm>I</fnm></au>
  </aug>
  <source>Nature Physics</source>
  <pubdate>2019</pubdate>
  <volume>15</volume>
  <issue>4</issue>
  <fpage>313</fpage>
  <lpage>-320</lpage>
</bibl>

<bibl id="B17">
  <title><p>Network Analysis of Ballast-Mediated Species Transfer Reveals
  Important Introduction and Dispersal Patterns in the {{Arctic}}</p></title>
  <aug>
    <au><snm>Saebi</snm><fnm>M</fnm></au>
    <au><snm>Xu</snm><fnm>J</fnm></au>
    <au><snm>Curasi</snm><fnm>SR</fnm></au>
    <au><snm>Grey</snm><fnm>EK</fnm></au>
    <au><snm>Chawla</snm><fnm>NV</fnm></au>
    <au><snm>Lodge</snm><fnm>DM</fnm></au>
  </aug>
  <source>Scientific Reports</source>
  <pubdate>2020</pubdate>
  <volume>10</volume>
  <issue>1</issue>
  <fpage>19558</fpage>
</bibl>

<bibl id="B18">
  <title><p>Representing Higher-Order Dependencies in Networks</p></title>
  <aug>
    <au><snm>Xu</snm><fnm>J</fnm></au>
    <au><snm>Wickramarathne</snm><fnm>TL</fnm></au>
    <au><snm>Chawla</snm><fnm>NV</fnm></au>
  </aug>
  <source>Science Advances</source>
  <pubdate>2016</pubdate>
  <volume>2</volume>
  <issue>5</issue>
  <fpage>e1600028</fpage>
  <lpage>e1600028</lpage>
</bibl>

<bibl id="B19">
  <title><p>Engineering {{Label}}-{{Constrained Shortest}}-{{Path
  Algorithms}}</p></title>
  <aug>
    <au><snm>Barrett</snm><fnm>C</fnm></au>
    <au><snm>Bisset</snm><fnm>K</fnm></au>
    <au><snm>Holzer</snm><fnm>M</fnm></au>
    <au><snm>Konjevod</snm><fnm>G</fnm></au>
    <au><snm>Marathe</snm><fnm>M</fnm></au>
    <au><snm>Wagner</snm><fnm>D</fnm></au>
  </aug>
  <source>Algorithmic {{Aspects}} in {{Information}} and
  {{Management}}</source>
  <editor>Fleischer, Rudolf and Xu, Jinhui</editor>
  <pubdate>2008</pubdate>
  <volume>5034</volume>
  <fpage>27</fpage>
  <lpage>-37</lpage>
</bibl>

<bibl id="B20">
  <title><p>Fast {{Routing}} in {{Very Large Public Transportation Networks
  Using Transfer Patterns}}</p></title>
  <aug>
    <au><snm>Bast</snm><fnm>H</fnm></au>
    <au><snm>Carlsson</snm><fnm>E</fnm></au>
    <au><snm>Eigenwillig</snm><fnm>A</fnm></au>
    <au><snm>Geisberger</snm><fnm>R</fnm></au>
    <au><snm>Harrelson</snm><fnm>C</fnm></au>
    <au><snm>Raychev</snm><fnm>V</fnm></au>
    <au><snm>Viger</snm><fnm>F</fnm></au>
  </aug>
  <source>Algorithms \textendash{} {{ESA}} 2010</source>
  <editor>Hutchison, David and Kanade, Takeo and Kittler, Josef and Kleinberg,
  Jon M. and Mattern, Friedemann and Mitchell, John C. and Naor, Moni and
  Nierstrasz, Oscar and Pandu Rangan, C. and Steffen, Bernhard and Sudan, Madhu
  and Terzopoulos, Demetri and Tygar, Doug and Vardi, Moshe Y. and Weikum,
  Gerhard and {de Berg}, Mark and Meyer, Ulrich</editor>
  <pubdate>2010</pubdate>
  <volume>6346</volume>
  <fpage>290</fpage>
  <lpage>-301</lpage>
</bibl>

<bibl id="B21">
  <title><p>Shortest Viable Path Algorithm in Multimodal Networks</p></title>
  <aug>
    <au><snm>Lozano</snm><fnm>A</fnm></au>
    <au><snm>Storchi</snm><fnm>G</fnm></au>
  </aug>
  <source>Transportation Research Part A: Policy and Practice</source>
  <pubdate>2001</pubdate>
  <volume>35</volume>
  <issue>3</issue>
  <fpage>225</fpage>
  <lpage>-241</lpage>
</bibl>

<bibl id="B22">
  <title><p>Algorithms for {{Finding Shortest Paths}} in {{Networks}} with
  {{Vertex Transfer Penalties}}</p></title>
  <aug>
    <au><snm>Lewis</snm><fnm>R</fnm></au>
  </aug>
  <source>Algorithms</source>
  <pubdate>2020</pubdate>
  <volume>13</volume>
  <issue>11</issue>
  <fpage>269</fpage>
</bibl>

<bibl id="B23">
  <title><p>The K-Color Shortest Path Problem</p></title>
  <aug>
    <au><snm>Ferone</snm><fnm>D</fnm></au>
    <au><snm>Festa</snm><fnm>P</fnm></au>
    <au><snm>Pastore</snm><fnm>T</fnm></au>
  </aug>
  <source>Advances in Optimization and Decision Science for Society, Services
  and Enterprises</source>
  <publisher>{Cham}: {Springer International Publishing}</publisher>
  <editor>Paolucci, Massimo and Sciomachen, Anna and Uberti, Pierpaolo</editor>
  <pubdate>2019</pubdate>
  <volume>3</volume>
  <fpage>367</fpage>
  <lpage>-376</lpage>
</bibl>

<bibl id="B24">
  <title><p>Computing and {{Listing}} St-{{Paths}} in {{Public Transportation
  Networks}}</p></title>
  <aug>
    <au><snm>B{\"o}hmov{\'a}</snm><fnm>K</fnm></au>
    <au><snm>H{\"a}fliger</snm><fnm>L</fnm></au>
    <au><snm>Mihal{\'a}k</snm><fnm>M</fnm></au>
    <au><snm>Pr{\"o}ger</snm><fnm>T</fnm></au>
    <au><snm>Sacomoto</snm><fnm>G</fnm></au>
    <au><snm>Sagot</snm><fnm>MF</fnm></au>
  </aug>
  <source>Theory of Computing Systems</source>
  <pubdate>2018</pubdate>
  <volume>62</volume>
  <issue>3</issue>
  <fpage>600</fpage>
  <lpage>-621</lpage>
</bibl>

<bibl id="B25">
  <title><p>HYPA: Efficient Detection of Path Anomalies in Time Series Data on
  Networks</p></title>
  <aug>
    <au><snm>LaRock</snm><fnm>T</fnm></au>
    <au><snm>Nanumyan</snm><fnm>V</fnm></au>
    <au><snm>Scholtes</snm><fnm>I</fnm></au>
    <au><snm>Casiraghi</snm><fnm>G</fnm></au>
    <au><snm>{Eliassi-Rad}</snm><fnm>T</fnm></au>
    <au><snm>Schweitzer</snm><fnm>F</fnm></au>
  </aug>
  <source>Proceedings of the 2020 SIAM International Conference on Data
  Mining</source>
  <pubdate>2020</pubdate>
  <fpage>460</fpage>
  <lpage>-468</lpage>
</bibl>

<bibl id="B26">
  <title><p>A universal distribution law of network detour ratios</p></title>
  <aug>
    <au><snm>Yang</snm><fnm>H</fnm></au>
    <au><snm>Ke</snm><fnm>J</fnm></au>
    <au><snm>Ye</snm><fnm>J</fnm></au>
  </aug>
  <source>Transportation Research Part C: Emerging Technologies</source>
  <pubdate>2018</pubdate>
  <volume>96</volume>
  <fpage>22</fpage>
  <lpage>-37</lpage>
</bibl>

<bibl id="B27">
  <title><p>Fast Unfolding of Communities in Large Networks</p></title>
  <aug>
    <au><snm>Blondel</snm><fnm>VD</fnm></au>
    <au><snm>Guillaume</snm><fnm>JL</fnm></au>
    <au><snm>Lambiotte</snm><fnm>R</fnm></au>
    <au><snm>Lefebvre</snm><fnm>E</fnm></au>
  </aug>
  <source>Journal of Statistical Mechanics: Theory and Experiment</source>
  <pubdate>2008</pubdate>
  <volume>2008</volume>
  <issue>10</issue>
  <fpage>P10008</fpage>
</bibl>

<bibl id="B28">
  <title><p>Rank Correlation Methods</p></title>
  <aug>
    <au><snm>Kendall</snm><fnm>MG</fnm></au>
  </aug>
  <publisher>{London}: {Griffin}</publisher>
  <edition>4</edition>
  <pubdate>1970</pubdate>
</bibl>

<bibl id="B29">
  <title><p>{{SciPy} 1.0: Fundamental Algorithms for Scientific Computing in
  Python}</p></title>
  <aug>
    <au><snm>Virtanen</snm><fnm>P</fnm></au>
    <au><snm>Gommers</snm><fnm>R</fnm></au>
    <au><snm>Oliphant</snm><fnm>TE</fnm></au>
    <au><snm>Haberland</snm><fnm>M</fnm></au>
    <au><snm>Reddy</snm><fnm>T</fnm></au>
    <au><snm>Cournapeau</snm><fnm>D</fnm></au>
    <au><snm>Burovski</snm><fnm>E</fnm></au>
    <au><snm>Peterson</snm><fnm>P</fnm></au>
    <au><snm>Weckesser</snm><fnm>W</fnm></au>
    <au><snm>Bright</snm><fnm>J</fnm></au>
    <au><snm>{van der Walt}</snm><fnm>SJ</fnm></au>
    <au><snm>Brett</snm><fnm>M</fnm></au>
    <au><snm>Wilson</snm><fnm>J</fnm></au>
    <au><snm>Millman</snm><fnm>KJ</fnm></au>
    <au><snm>Mayorov</snm><fnm>N</fnm></au>
    <au><snm>Nelson</snm><fnm>ARJ</fnm></au>
    <au><snm>Jones</snm><fnm>E</fnm></au>
    <au><snm>Kern</snm><fnm>R</fnm></au>
    <au><snm>Larson</snm><fnm>E</fnm></au>
    <au><snm>Carey</snm><fnm>C J</fnm></au>
    <au><snm>Polat</snm><fnm>{\.I}</fnm></au>
    <au><snm>Feng</snm><fnm>Y</fnm></au>
    <au><snm>Moore</snm><fnm>EW</fnm></au>
    <au><snm>{VanderPlas}</snm><fnm>J</fnm></au>
    <au><snm>Laxalde</snm><fnm>D</fnm></au>
    <au><snm>Perktold</snm><fnm>J</fnm></au>
    <au><snm>Cimrman</snm><fnm>R</fnm></au>
    <au><snm>Henriksen</snm><fnm>I</fnm></au>
    <au><snm>Quintero</snm><fnm>E. A.</fnm></au>
    <au><snm>Harris</snm><fnm>CR</fnm></au>
    <au><snm>Archibald</snm><fnm>AM</fnm></au>
    <au><snm>Ribeiro</snm><fnm>AH</fnm></au>
    <au><snm>Pedregosa</snm><fnm>F</fnm></au>
    <au><snm>{van Mulbregt}</snm><fnm>P</fnm></au>
    <au><cnm>{SciPy 1.0 Contributors}</cnm></au>
  </aug>
  <source>Nature Methods</source>
  <pubdate>2020</pubdate>
  <volume>17</volume>
  <fpage>261</fpage>
  <lpage>-272</lpage>
</bibl>

<bibl id="B30">
  <title><p>Higher-Order Aggregate Networks in the Analysis of Temporal
  Networks: Path Structures and Centralities</p></title>
  <aug>
    <au><snm>Scholtes</snm><fnm>I</fnm></au>
    <au><snm>Wider</snm><fnm>N</fnm></au>
    <au><snm>Garas</snm><fnm>A</fnm></au>
  </aug>
  <source>The European Physical Journal B</source>
  <pubdate>2016</pubdate>
  <volume>89</volume>
  <issue>3</issue>
</bibl>

</refgrp>
} 
		
		
		
		
		\appendix
	\end{backmatter}
	\newpage
	\section{Comparison of Structural Core Results for Varying Thresholds}
	\label{app:core}
	
\begin{figure}[ht!]
		\centering
		\includegraphics[scale=0.83]{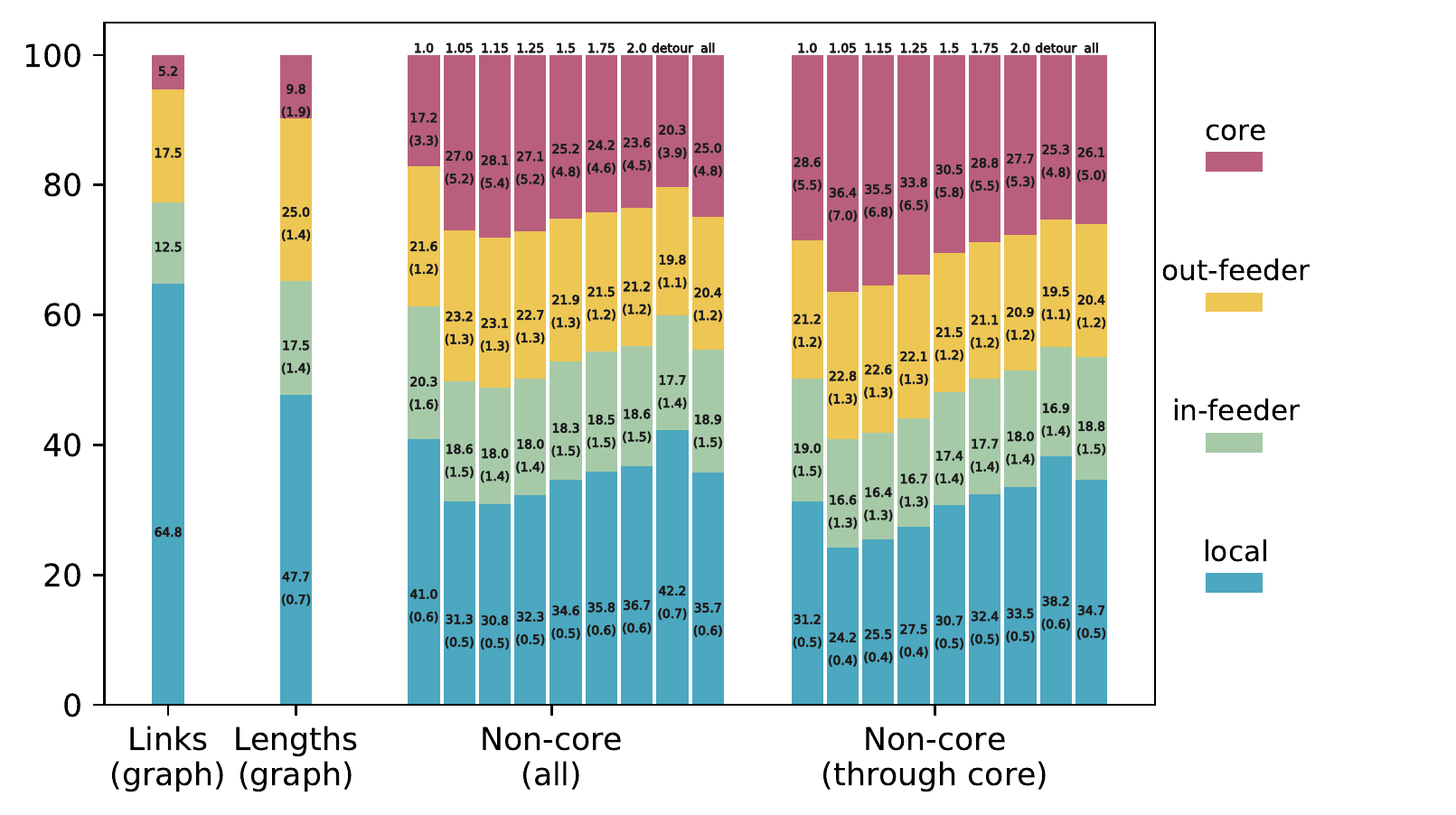}
		\caption{Percentage of links and shipping length by edge type for various values of distance threshold $\alpha$, the detour factor threshold ("detour"), as well as using the entire set of minimum-route paths ("all"). As in Figure~\ref{fig:linkandlengthpct}, numbers in parentheses correspond to length-to-link percentage ratios. Results are similar for paths between all non-core ports and paths between non-core ports that include at least one core link, and results are broadly similar across threshold values. Looking at all of the paths between non-core ports, when only the minimum shipping distance path is kept ($\alpha=1.0$), core links account for about 17.1\% of the total length, but after adding a small number of paths close to the minimum distance ($\alpha=1.05$) the involvement of the core increases to 26.8\%. As $\alpha$ increases, the percentage of core links varies between 23.6\% and 28.1\%, while the detour threshold percentage is 20.3\%. Turning to the paths between non-core ports that pass through the core, at $\alpha=1.0$ core links account for about 28.5\% of the total length, but at $\alpha=1.05$ the involvement of the core increases to 35.9\%. As $\alpha$ grows and fewer paths are filtered, the length percentage accounted for by the core decreases to 26.1\% when all minimum-route paths are included, and is 25.3\% using detour factor filtering.}
		\label{fig:linkandlengthpct-app}
	\end{figure}

	In the main text we showed the link and length percentages for minimum-route paths using the distance filter $\alpha=1.15$. In Figure~\ref{fig:linkandlengthpct-app}, we present results using all thresholding schemes, including detour factor thresholding ("detour") and no filtering ("all"). Regardless of the extent of filtering, we find the same result: the statistics of core links were simultaneously underestimated and overestimated in previous work, while the statistics of local links were underestimated; for detailed illustration, refer to Figure~\ref{fig:linkandlengthpct} and its associated main text.
	
	\section{Relationship Between Number of Paths and Path Length}
	\label{app:mVp}
	
\begin{figure}[ht!]
		\centering
		\includegraphics[width=0.9\textwidth]{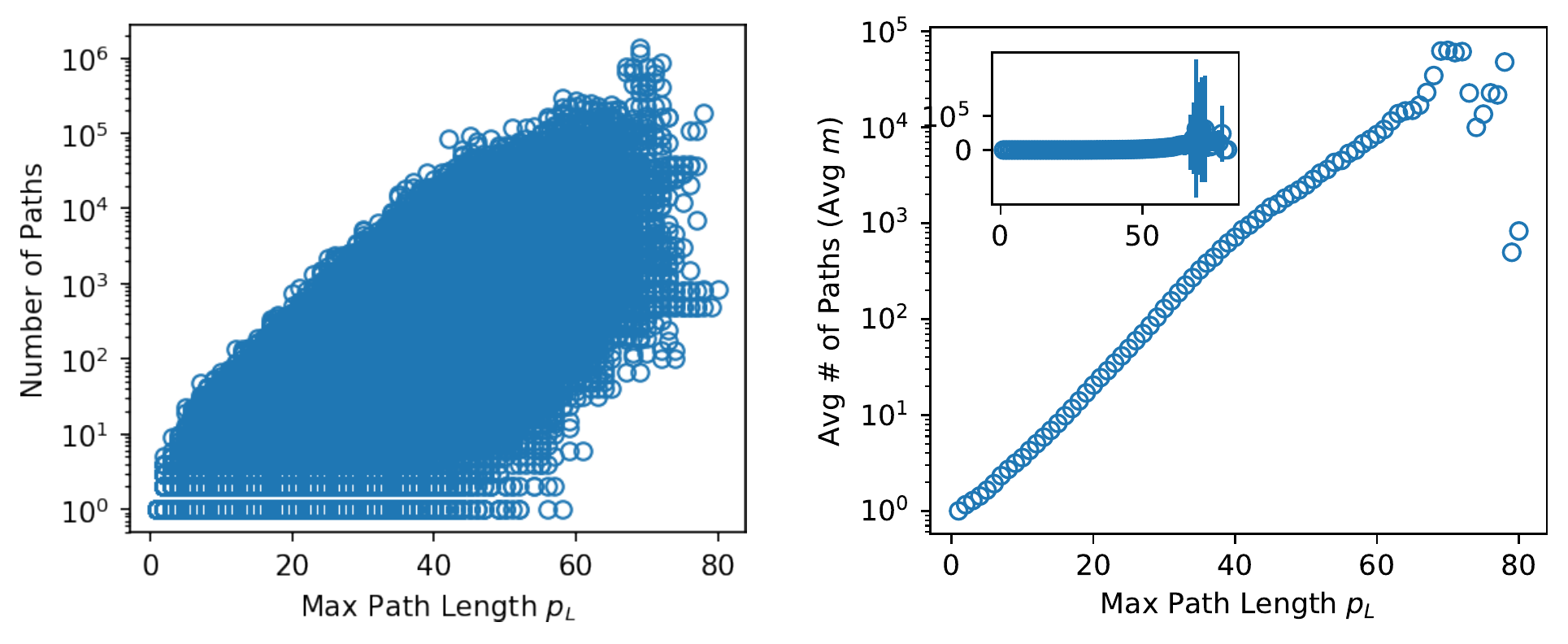}
		\caption{Maximum path length against the log of the number of paths per pair of source and target ports (left) as well as the log of the average number of paths (right, main plot) and including the amount of variation measured by one standard deviation (right, inset plot). As the maximum path length between a pair increases, so does the average number of paths between pairs with that maximum path length. An intuitive explanation for this trend is that longer minimum-route paths are concatenations of shorter minimum-route paths between intermediate source and target pairs (e.g. the loop on line 12 of Algorithm~\ref{alg:minroutes-iterative}) that are all interchangeable. This implies that the number of paths between a pair with large maximum path length is a function of the product of the number of minimum-route paths at smaller lengths, and so the number of paths grows much more quickly than the maximum path length. However, at the highest maximum path lengths this trend does not hold. We attribute this to the fact that the very longest paths are likely to appear between ports that are not well-connected, thus there are few (perhaps only 1) viable minimum-route paths between some of these pairs, dragging the average down. Put another way: very long paths usually occur between poorly connected nodes, meaning they are more likely to be unique or few in number.}
		\label{fig:mVp-app}
	\end{figure}

	In Figure~\ref{fig:mVp-app} we show the relationship between the number of minimum-route paths and the maximum length among those paths for all pairs of ports. In the left plot we plot these quantities for every pair, while in the right plot we show the average number of paths for each length, with error bars shown in the inset plot. As the maximum path length increases, the number of paths per pair also increases, but at a much faster rate. This is evidence that $m$ (number of paths) dominates $p_L$ (maximum path length) in the runtime calculation for filtering minimum-route paths in Section~\ref{subsec:filt-runtime}.
	
\end{document}